\documentclass[a4paper,12pt]{article}
\pdfoutput=1 
\usepackage{jheppub} 
\usepackage{amsmath,amssymb}
\usepackage{graphicx}
\def\dhalf{\frac{d}{2}}

\def\Pbox{{\rm P}_{\rm box}}
\def\Xbox{{\rm X}_{\rm box}}
\def\Dbox{{\rm D}_{\rm box}}
\def\DboxM{{\overline {\rm D}}_{\rm box}}
\def\Bbox{{\rm B}_{\rm box}}

\def\bea{\begin{eqnarray}}
\def\eea{\end{eqnarray}}

\title{Removing infrared divergences from two-loop integrals}

\author[a]{Charalampos Anastasiou}
\author[b]{George Sterman}
\affiliation[a]{Institute for Theoretical Physics, ETH Zurich, 8093 Z\"urich, Switzerland}
\affiliation[b]{C.N. Yang Institute for Theoretical Physics and Department of Physics and Astronomy, Stony Brook University, Stony Brook, NY 11794,
USA}
\emailAdd{babis@phys.ethz.ch}
\emailAdd{george.sterman@stonybrook.edu}

\abstract{ 
Feynman amplitudes at higher orders in perturbation theory generically
have complex singular structures.  Notwithstanding the emergence of many
powerful new methods, the presence of infrared divergences poses
significant challenges for their evaluation.  In this article, we
develop a systematic method for the removal of the infrared
singularities, by adding appropriate counterterms that approximate and
cancel divergent limits point-by-point at the level of the integrand.  We
provide a proof of concept for our method  by  applying it to
master-integrals that are found in scattering amplitudes for 
representative $2 \to 2$ scattering processes of massless particles. We
demonstrate that, after the  introduction of counterterms,  the
remainder 
is finite in four dimensions.  In
addition, we find in these cases that the complete singular dependence of the integrals can be
obtained simply by analytically integrating the counterterms.  
Finally, we observe that our subtraction method can be also 
useful in order to extract in a simple way the asymptotic behavior of 
Feynman amplitudes in the limit of small mass parameters. 
}


\begin{document}

\maketitle 
\flushbottom

\newpage
\section{Introduction}
\label{sec:introduction}

The drive for precision in collider cross sections has become a major theme in contemporary high energy physics.   Precision requires elements from theory, experiment and data analysis, and a major requirement in theory involves the calculation, including numerical evaluation, of multi-loop perturbative amplitudes and multi-particle integrals over restricted regions of phase space.   As a practical matter, it is often the case that amplitudes require infrared regularization, and unphysical contributions  that must cancel in physical quantities.   In addition, the presence of massive particles in multi-loop amplitudes in combination with massless lines, while simplifying the overall infrared structure, often results in an even more difficult burden for analytic integrations.   Similarly, even at fixed loops, large numbers of external massless lines leads to further complexity.   

With all this in mind, it should be helpful to develop additional methods to separate systematically, and if possible algorithmically, infrared-divergent integration regions from infrared finite, treating the latter numerically and the former, analytically.  Such methods have proved invaluable at next-to-leading order~\cite{Giele:1993dj,Frixione:1995ms,Catani:1996vz}, and considerable progress has already been made at NNLO~\cite{Anastasiou:2003gr,GehrmannDeRidder:2005cm,Daleo:2006xa, Somogyi:2006da,DelDuca:2016ily, Catani:2007vq,Czakon:2010td,Boughezal:2013uia,Currie:2013vh,Boughezal:2015dva,Gaunt:2015pea,Cacciari:2015jma,Caola:2017dug,Currie:2018oxh,Herzog:2018ily,Grazzini:2017mhc,Boughezal:2018mvf,Behring:2018cvx,Czakon:2018iev} and beyond~\cite{Mistlberger:2018etf,Dreyer:2016oyx,Dulat:2018bfe,Cieri:2018oms,Currie:2018fgr,Dreyer:2018qbw,Ruijl:2018poj,Chetyrkin:2018dce}.   In the following, we explore the use of methods inspired by those that have been developed to prove the all-orders factorization of amplitudes and cross sections to a number of examples.  Our goal is to provide an in principle demonstration that such an approach, based on the nested local subtractions of infrared integration regions, may be flexible, practical and useful.   Although we will restrict ourselves to amplitudes in this discussion, we hope the method will provide a way toward the combination of virtual and real corrections for infrared safe quantities without the need for infrared regularization.

Nested subtractions play a central role in the proof of factorization for inclusive color-singlet hard-scattering cross sections, such as Drell-Yan production \cite{Collins:2011zzd}, and have been developed for a proof of factorization for fixed-angle scattering in gauge theories \cite{Erdogan:2014gha}.   The value of factorization in organizing the calculation of higher-order cross sections has been explored in Refs.\ \cite{Magnea:2018ebr,Magnea:2018hab}.   Here, we will concentrate on amplitudes, and illustrate the feasibility of the direct evaluation of two loop diagrams for fixed angle scattering, based on our general knowledge of infrared structure.

Our approach will be somewhat distinct from methods used in the proof of factorization in gauge theory cross sections and amplitudes.   We will start with purely scalar diagrams in $\phi^3$ theory, and indeed, we will find that some of the severe infrared behavior found in even simple scalar examples can be treated with a straightforward approach.   

In the following section we recall the infrared structure of fixed-angle scattering, describe the approach we will take, based on ordered subtractions, and illustrate the method as it applies to the one-loop box diagram.   In Sec.\ \ref{sec:examples} we show how this method operates when applied to essential two-loop examples, including the double box and crossed double box.  Finally, in Sec.\ \ref{sec:smallmass} we discuss the use of the method in generating asymptotic expansions in small masses that regulate infrared divergences, replacing them with logarithmic dependence.    


\section{Method for subtracting the infrared divergences in loop integrals}  
\label{sec:method}

In this paper, we treat scalar amplitudes that describe fixed-angle scattering, reorganizing them into a sum of terms, each of which is an infrared-sensitive integral that can be performed analytically, multiplied by a coefficient integral that 
is a tree diagram or which
can be computed numerically.   To these terms will be added a ``remainder", which is also free of infrared divergence. In particular, we can imagine evaluating the coefficients directly in four dimensions.   It is worth noting that in dimensional regularization, it would be necessary to expand these coefficients in $\epsilon=2-d/2$ around $\epsilon=0$ in $d$ dimensions to derive the full expression for the amplitude, but we anticipate that in the calculation of infrared-safe cross sections, this will not be necessary.   

\subsection{Leading regions and power counting}

\begin{figure}[h]
\begin{center}
\includegraphics[width=0.4\textwidth]{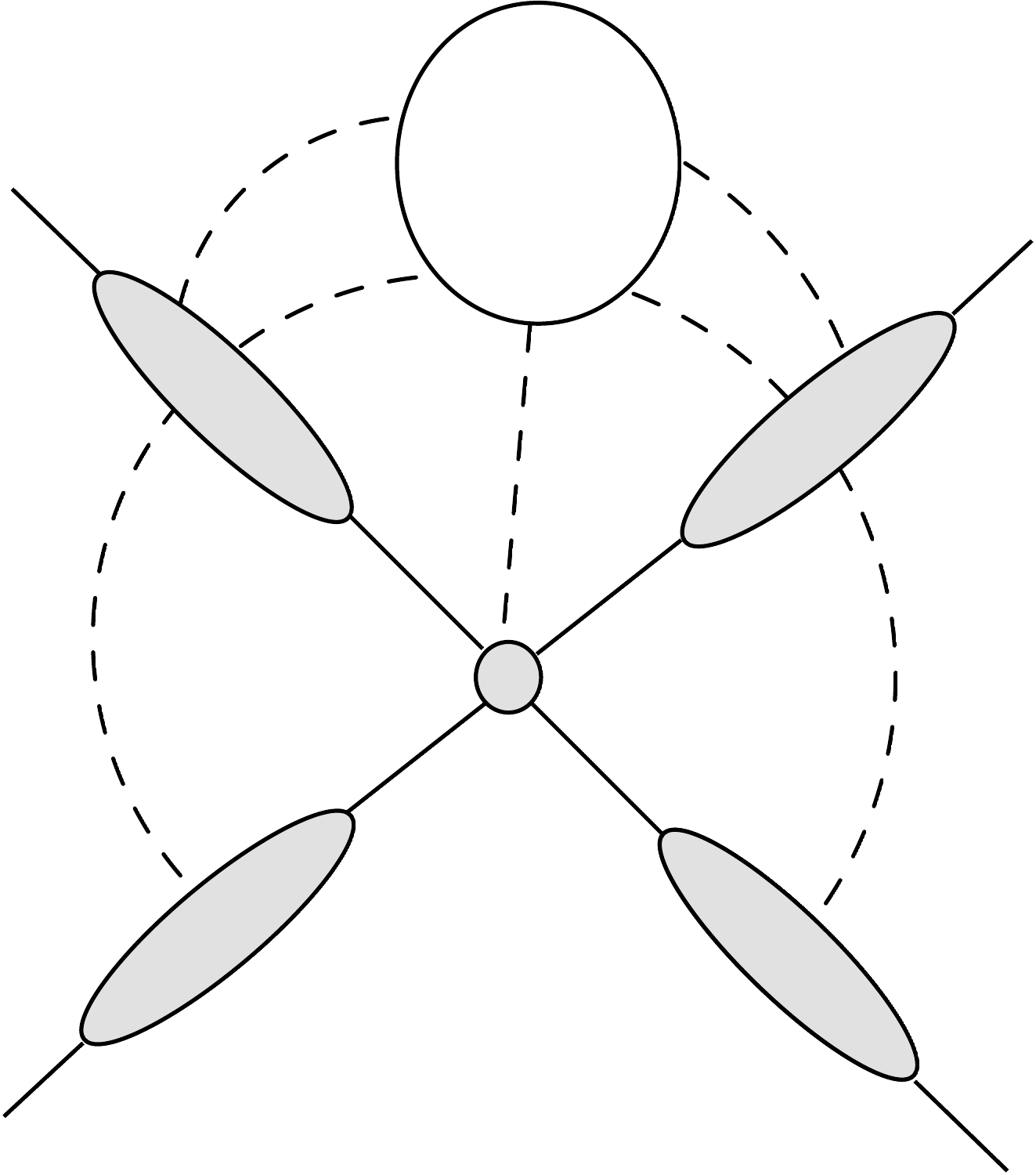}
\caption{\label{fig:pinch surface figure} Depiction of a general pinch surface for two-to-two scattering.  Shaded blobs represent jet subdiagrams, and the open circle a subdiagram of ``soft" lines, whose momenta vanish at the pinch surface.  In the center is a ``hard" subdiagram consisting of lines off-shell by the order of the momentum transfer. Each line connecting the soft, jet and hard subdiagrams represents an arbitrary number of lines.  For the purposes of this discussion, all lines are scalars, and the dashed lines simply represent soft lines attached to jet subdiagrams.  Note the possibility that soft lines as well as jet lines attach to the hard subdiagram.}
\end{center}
\end{figure}

The essential observation that makes the subtraction method possible is that the sources of infrared divergence in fixed-angle scattering amplitudes are associated with a finite list of ``leading regions" in loop momentum space, which can be enumerated for an arbitrary diagram.   The generic form follows from the application of the general Coleman-Norton  criterion \cite{Coleman:1965xm} for a Landau pinch  surface \cite{Landau:1959fi,Sterman:1978bi} to fixed angle scattering \cite{Sen:1982bt,Sterman:2002qn}.   In this case, pinch surfaces are associated with configurations in which internal lines are either part of a ``jet subdiagram", of on-shell lines connected to one of the external lines, with all line momenta parallel to that line, or 
are in the ``soft subdiagram", and
have vanishing momentum.  All other lines are off-shell.   The general case is illustrated for two-to-two scattering in Fig.\ \ref{fig:pinch surface figure}.   

In general, lines with vanishing momentum (soft lines) may be attached either to each other in the soft subdiagram or to the jet functions.   In gauge theories, pinch surfaces at which soft lines connect to the hard subdiagram are power-counting suppressed, but in pure scalar theories this is not always the case.    

The behavior of integrals in the neighborhood of an arbitrary pinch surface is particularly easy to determine in scalar theories.  As illustrated in the figure, a generic pinch surface, labelled $\gamma$, can be characterized by some number of collinear loop momenta, $L_C^\gamma$, and by $L^\gamma_S$ soft loop momenta.   Correspondingly, at pinch surface $\gamma$, there are $N_S^\gamma$ ``soft" lines, with vanishing momentum, and $N_C$ ``jet" lines, each of whose momentum is a fraction $x$, with $0<x<1$ of the momentum of one of the external lines.   To find the behavior of the integral at pinch surface $\gamma$, we introduce a dimensionless scaling variable $\delta$ and study the behavior of the integrand and integration volume for $\delta \rightarrow 0$, where it takes all momenta to the pinch surface. To keep track of dimensions, we label by $Q$ the typical hard-scattering momentum scale, say $Q\sim \sqrt{\hat s}$, for $2\rightarrow n$ fixed angle scaling.   

Now, for soft lines, which vanish in all four components at the pinch surface, we take
\bea
\label{eq:soft-scale}
k_i^\mu\ \sim \ \delta Q\, .
\eea
Jet line momenta, on the other hand, approach a fraction of the corresponding external momentum according to
\bea
\label{eq:collinear_momentum}
k_j^\mu\ &\sim x_j\, p^\mu\ + \beta_j \eta_p^\mu  + \   k_\perp^\mu\, ,
\eea
where $\eta_p^\mu$ is a lightlike vector moving opposite to $p^\mu$, with $\eta_p^2=0$, and where $p\cdot k_\perp =\eta_p\cdot k_\perp=0$.   The scalings for these jet line components  are then
\bea
\label{eq:collinear-scaling}
\beta_j \ &\sim&\ \delta\, Q\, ,
\nonumber\\
k_\perp\ &\sim&\ \delta^{1/2}\, Q\, .
\eea
With all $x_i$ fixed, the integrand times volume element of the remaining integrals in $d$ dimensions then behaves as $\delta^{p_\gamma}$, with $p_\gamma$ a ``degree of divergence", given in terms of the number of loops and lines by
\bea
p_\gamma\ =\ dL_S^\gamma\ +\ 2L_C^\gamma\ -\ 2N_S\ -\ N_C\, .
\label{eq:power-ctg-scalar}
\eea
For $p^\gamma>0$, the integral is finite near the pinch surface.   When $p_\gamma=0$ there is a logarithmic divergence, and when $p_\gamma$ is negative, a power divergence.   We note that a generic pinch surface is contained in surfaces of larger dimensionality, in which some subset of loop momenta are finite distances away from these configurations, while others remain close to them.  A straightforward analysis of these integrals shows that for all such surfaces, the appropriate scaling follows the same rule \cite{Sterman:1978bi}.   An essential feature is that the pinch surfaces of the integrals that result by keeping only the leading behavior for a given pinch surface itself has (lower-dimensional) pinch surfaces and leading regions that are subsets of the pinch surfaces of the original diagram \cite{Sterman:1994ce}.    It should be noted that this is a feature of fixed-angle scattering, and that in other configurations, in particular for forward scattering, additional power counting analysis is necessary.

For a massless cubic scalar theory, the degree of divergence can be arbitrarily negative at high orders, depending on the diagram.   This power infrared behavior is the result of the dimensional (superrenormalizable) cubic coupling.   In the cases below, for two-to-two fixed angle scattering, we will in fact encounter both logarithmic and power singularities.   Although it is not our intention to propose an all-orders treatment of the scalar case, our method will deal with the power as well as logarithmic singularities.
  
As noted, leading regions are connected in general to other leading regions, and while some are contained within others, others may be disjoint or overlapping.   The situation is analogous to the classification of ultraviolet divergences, in which divergent subdiagrams may nest or may overlap with each other.   Our aim here is to show how, given any $L$-loop diagram defined in $d=4-2\epsilon$ dimensions, 
\bea
I^{(L)}(\epsilon) \ =\ \int \prod_{i=1}^L\, \frac{d^dk_i}{i\pi^{d/2}}\, f(k_i)\, ,
\eea
we can construct an integral that is finite in four dimensions, by a suitable subtraction,
\bea
I^{(L)}(\epsilon)\ &=&\ \int \prod_{i=1}^L\, \frac{d^dk_i}{i\pi^{d/2}}\ \left [ f(k_i)\ -\ f_{\rm approx}(k_i) \right ]\ 
+ \int \prod_{i=1}^L\, \frac{d^dk_i}{i\pi^{d/2}}\  f_{\rm approx}(k_i) 
\nonumber\\
&=& I_{\rm finite}^{(L)}(\epsilon)\ +\ \int \prod_{i=1}^L\, \frac{d^dk_i}{i\pi^{d/2}}\ f_{\rm approx}(k_i) .
\label{eq:f-k-subtract}
\eea
where $I_{\rm finite}(\epsilon)$ has a finite $\epsilon\rightarrow 0$ limit.

The essential result of perturbative ultraviolet renormalization is that sums of products of multiple, nested subtractions produce finite Green functions.   It is not necessary to make sequential subtractions involving overlapping regions.  A similar structure has been developed in infrared subtraction formalisms, starting as early as Ref.\ \cite{Collins:1981uk}.  Following the notation of that paper, we can represent the result as
\bea
f_{\rm approx}(k_i)\ =\ \sum_{{\cal N}} \prod_{a \in {\cal N}}\ (-\, t_a)\ f(k_i)\, ,
\label{eq:forest}
\eea
where each product is over a non-empty, ordered set $\cal N$ of approximation operators $t_a$ associated with pinch surfaces $a$, which act to the right.  In Refs.\ \cite{Collins:2011zzd} and \cite{Erdogan:2014gha}, it was shown that sums of nested subtractions, starting from the smallest, most singular regions, can be used to separate infrared singularities from short-distance structure.   We shall not review the details of these arguments, but only observe that the pattern starts by making subtractions that match the behavior of the integrand in the most singular regions of momentum space, of the smallest volume, in which the largest numbers of lines approach the light cone (or more generally, the mass shell).   The nested operations then act systematically on the resulting terms to remove remaining divergences, by proceeding to subtract  the next largest volume, then the next, and so on.
This is possible because, as noted above, for fixed-angle scattering the pinch surfaces of the integrals after the action of the approximation operators are subsets of those of the original diagram.

For proofs of factorization in gauge theories, the approximations are tailored to match leading behavior, and often at the same time to provide expressions to which the Ward identities of the theory may be applied.  Generally, this results in the introduction of new ultraviolet divergences in subtractions, a feature that serves as a basis of resummation \cite{Kidonakis:1998nf}.  In our examples below, however, we set these considerations aside, and take a pragmatic approach to the identification of subtractions.   In particular, at this stage we design subtractions to avoid induced ultraviolet divergences.  Here a method introduced at one loop in Ref.\ \cite{Nagy:2003qn} will turn out to be useful.   This will already be apparent in our first example, the one-loop box diagram, to which we turn as a warm-up exercise in the following subsection.

\subsection{Subtraction for the one-loop box}
\label{sec:example_1loopbox}

\begin{figure}[h]
\begin{center}
\includegraphics[width=0.4\textwidth]{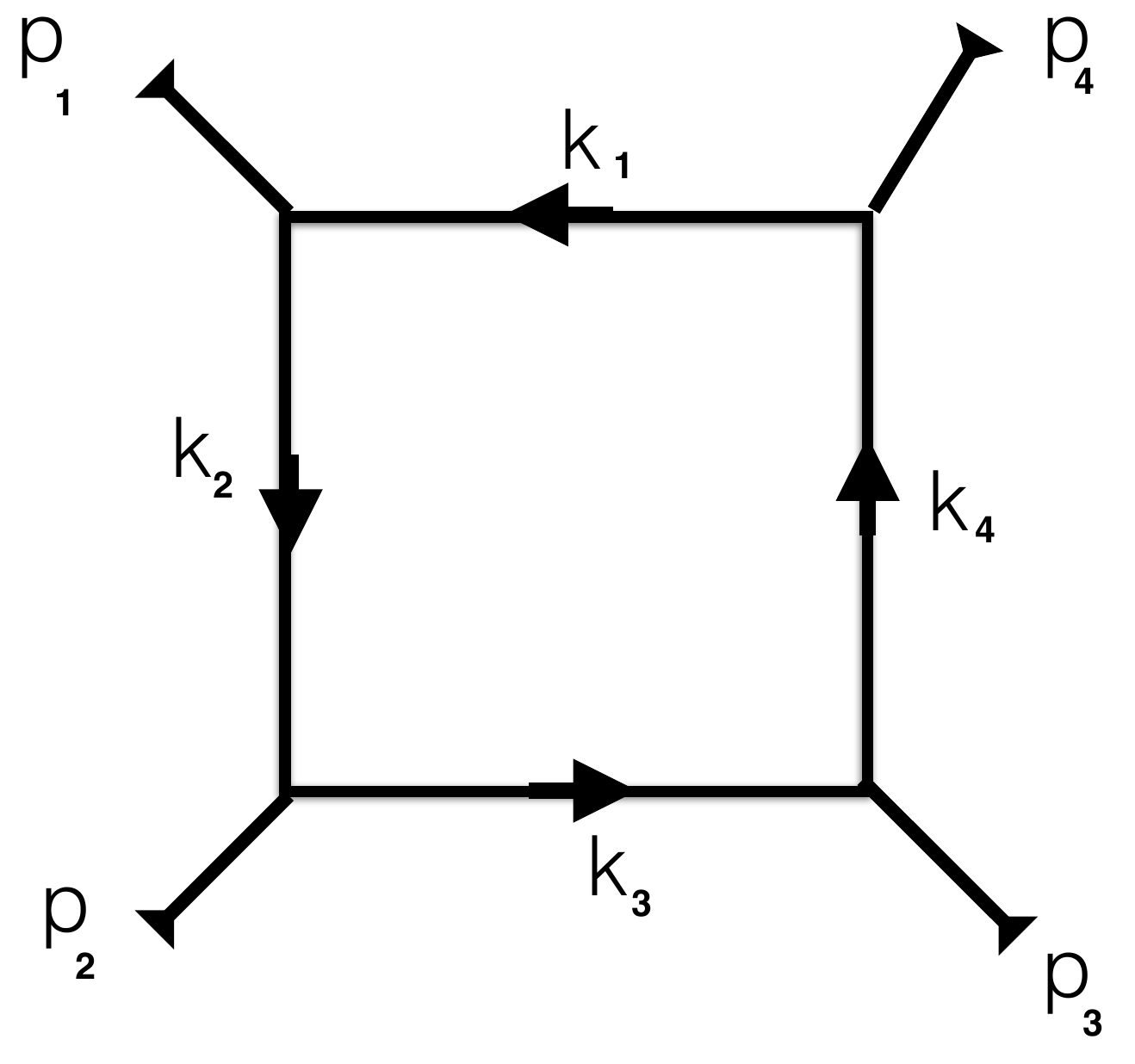}
\caption{\label{fig:box} The one-loop box}
\end{center}
\end{figure}

As a pedagogical example, we will apply the method of nested
subtractions to the massless one-loop scalar box integral, shown in Fig.~\ref{fig:box}.  We write the integral as  
\begin{equation}
\label{eq:1loopbox_def}
{\rm Box}  \equiv \int \frac{d^dk_1}{i \pi^{\frac d 2}} \frac{1}{A_1 A_2 A_3 A_4},
\end{equation}
where the propagator denominators are $A_j \equiv k_j^2+i0 , j=1\ldots 4$, and where
the internal momenta are related by
\begin{equation}
k_{j+1}=k_j +p_j\, , 
\end{equation}
with $k_4\equiv k_0$ in this notation.   The external momenta are taken all incoming, and satisfy
\begin{eqnarray}
p_i^2 =0, \quad p_{12}^2 \equiv (p_1+p_2)^2
&=& s, \quad p_{23}^2\equiv (p_2+p_3)^2=t,  
\nonumber\\
 p_{1234}\ \equiv\ p_1+p_2+p_3+p_4 &=&0,
\end{eqnarray}
 with $s,t$  two independent Mandelstam variables.  In the
 following, we will often use the shorthand notation $X_{ijk\ldots}
 =X_i+X_j+X_k + \ldots$. 

\begin{figure}[h]
\begin{center}
\includegraphics[width=0.8\textwidth]{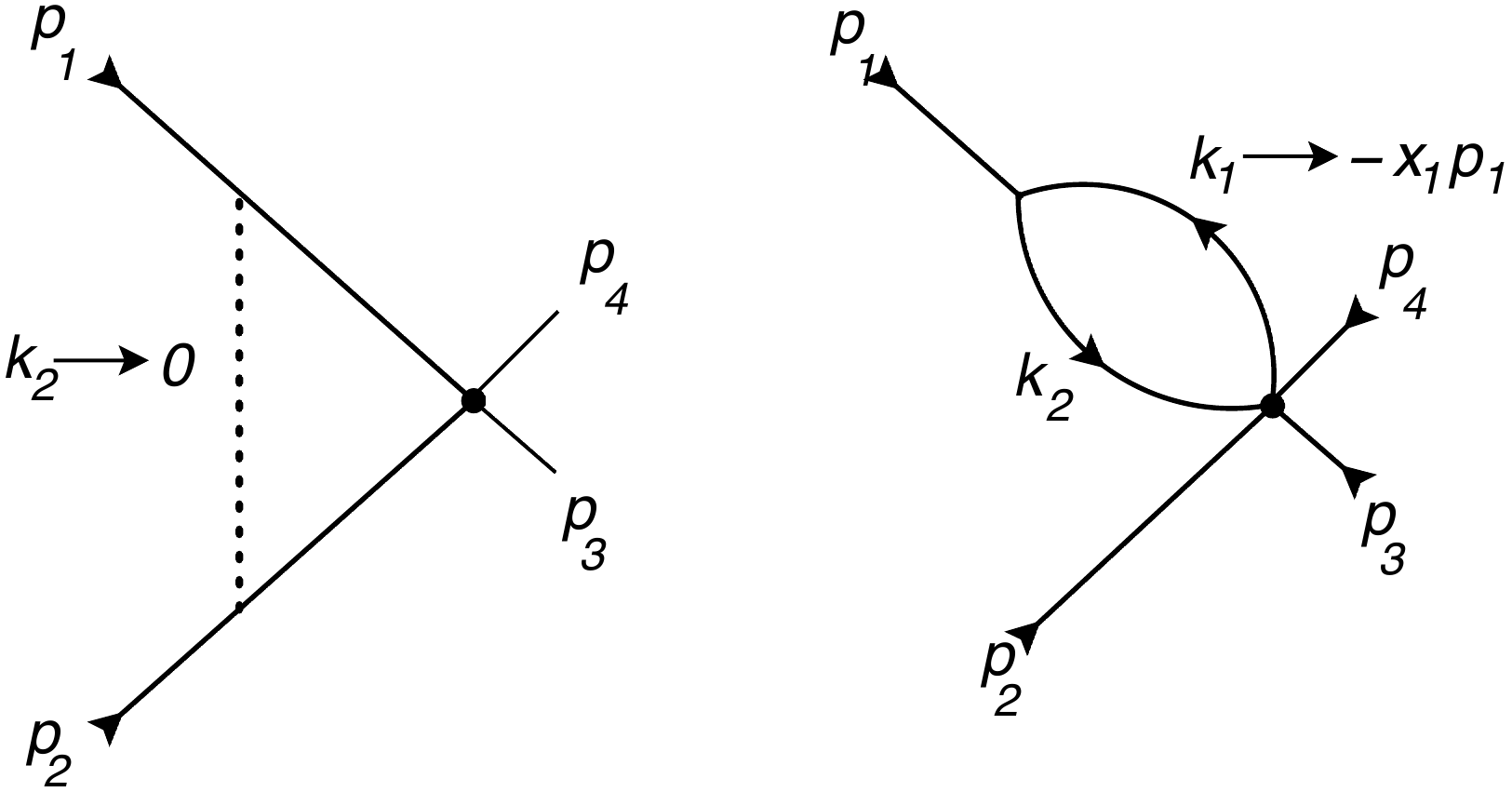}\\
 (a) \hspace{65mm} (b)
\caption{\label{fig:contracted_one_loop} Reduced diagrams for representative pinch surfaces of the one-loop box: (a)  soft limit $k_2\rightarrow 0$, (b) Collinear surface $k_1 \rightarrow -x_1p_1$.}
\end{center}
 \end{figure}

The integral of Eq.~\ref{eq:1loopbox_def} has infrared divergences, which
fall into the classes of leading regions identified in the previous section.
These leading regions are conventionally represented by ``reduced diagrams", in which
lines that are off-shell at the pinch surface are contracted to points.   
The eight leading pinch surfaces of the one-loop box fall into two categories,
illustrated by the examples of Fig.\ \ref{fig:contracted_one_loop}a and b.

First, the box is divergent in the four soft limits $k_i \sim \delta \to
0$, for which the leading regions are four disjoint points in loop momentum space, illustrated by Fig.\ \ref{fig:contracted_one_loop}a. 
In terms of the power counting of Eq.\ (\ref{eq:power-ctg-scalar}) these regions all have $L_S=1$, $N_S=1$, $N_C=2$,
corresponding to logarithmic divergence.   
Near the point $k_2=0$, for example, we have (Eq.\ (\ref{eq:soft-scale})),
$k_2^\mu \sim \delta \to 0$, and the denominators scale as 
\begin{eqnarray}
A_2 &=& k_2^2 \sim {\cal O}(\delta^2), 
\nonumber \\ 
A_1 &=&(k_2 -p_1)^2 \sim - 2 k_2 \cdot p_1 \sim
{\cal O}(\delta)\, , 
\nonumber\\
A_3 &=&(k_2 +p_2)^2 \sim  2 k_2 \cdot p_2 \sim
{\cal O}(\delta)\, , 
\nonumber \\
A_4 &=& (k_2 +p_{23})^2 \sim t + {\cal O}(\delta)\, .
\end{eqnarray}
We confirm that the integrand tends to 
\begin{equation}
\label{eq:box_k2_soft_strict}
\frac{d^d k_2}{A_1 A_2 A_3 A_4} \to \frac{d^d k_2}{(-2 k_2 \cdot p_1)
  k_2^2 (2 k_2 \cdot p_2) t} \sim {\cal O}(\delta^{d-4}).
\end{equation}
which is of course consistent with Eq.\ (\ref{eq:power-ctg-scalar}).

The integral of Eq.~(\ref{eq:1loopbox_def}) is also
divergent in the four collinear limits 
\begin{equation}
k_i \to -x_i p_i\, ,
\end{equation} 
for $i=1 \dots 4$ (note the directions of the arrows in Fig.\ \ref{fig:box}.)
For example, when $k_1,k_2$ become collinear to $p_1$,  illustrated by Fig.\ \ref{fig:contracted_one_loop}b, and
using the notation of Eq.\ (\ref{eq:collinear_momentum}), the loop momentum components in
\begin{equation}
\label{eq:a-collinear-limit}
k_1 = x_ 1 p_1 + \beta_1 \eta_1 + k_{1\perp}, 
\quad x_1 \equiv \frac{2 k_1 \cdot \eta_1}{ 2 p_1 \cdot \eta_1},  
\quad \beta_1 \equiv \frac{2 k_1 \cdot p_1}{ 2 p_1 \cdot \eta_1},   
\quad \eta_i^2=0, \quad \eta_i \cdot p_i \neq 0, 
\end{equation}
scale as in Eq.\ (\ref{eq:collinear_momentum}),
\begin{equation}
x_1 \sim {\cal O}(1), \quad \beta_1  \sim {\cal O}(\delta), \quad k_{1
  \perp} \sim {\cal O}(\sqrt \delta )   \, .
\end{equation}
In this region,
\begin{equation}
\label{eq:box_k2_collinear}
\frac{d^d k_2}{A_1 A_2 A_3 A_4} \to \frac{d^d k_2}{A_1
  A_2 s t x_1 (1-x_1)} \sim {\cal O}(\delta^{\frac{d}{2}-2}).
\end{equation}
Again, the power counting of Eq.\ (\ref{eq:power-ctg-scalar})  
indicates logarithmic divergence in four dimensions.   

For comparison below,
we give here the dimensionally-regulated expression for the one-loop box in $d=4-2\epsilon$ dimensions,
\bea
\label{eq:box-function}
{\rm Box}\ (s,t,\epsilon)\ &=& \frac{1}{st} \left\{ 
\frac{2 c_\Gamma}{\epsilon^2} 
\left[ (-s)^{-\epsilon} +(-t)^{-\epsilon} \right] - \pi^2 -\ln^2\left(
  \frac t s \right)
\right\} + {\cal O}(\epsilon),
\eea 
where 
\begin{equation}
c_\Gamma \equiv \frac{\Gamma^2(1-\epsilon) \Gamma(1+\epsilon)}{\Gamma(1-2\epsilon)}.
\end{equation}
We note the familiar double and single poles in $\epsilon$.  Our aim in this preliminary discussion is to 
see in practice how these poles are reproduced systematically in a subtraction formalism for this
simple case.

The method of nested subtractions introduces counterterms to the
integrand designed to remove the four soft and four collinear divergences of
the one-loop scalar box.  
The method removes first  
singularities from the smallest regions of the integration 
domain, and proceeds successively to remove the singularities in larger
volumes.   The regions of the soft singularities are clearly the smallest, since
they correspond to points in the integration domain ($k_i=0, i=1\ldots
4$) and so they will be removed first. 
In a soft limit, three of the propagators of the
one-loop box are on-shell and one
propagator is hard. The collinear singularities extend to larger
regions  $k_i = -x_i p_i, 0 < x_i\le 1 $ and, 
in the method of nested subtractions, they ought to be removed next.  In
the collinear limits, two propagators are on-shell and two are hard.   
We note as well that each soft region is an end-point of two  collinear regions.

We remove the divergence of the integral in the 
$k_2 \to 0$ limit by
subtracting a function that approximates the singular behavior of 
the integrand in that
limit.  We will sometimes refer to this subtraction as a counterterm. 
We may think of any such counterterm as the result of one approximation
operator in a product of subtractions, as illustrated in Eq.\ (\ref{eq:forest}). 
 Each particular approximating operation acts to produce a new integral, which approaches a singular expression like
Eq.~(\ref{eq:box_k2_soft_strict}) as $\delta \to 0$.  For some purposes,
in particular in proofs of factorization, a choice in which we keep only the terms with leading
behavior as $\delta\rightarrow 0$  is most convenient.   To be specific, let us label the subtraction operator
for the $k_2\rightarrow 0$ as $t_{S_2}$.
  This operator  acts as
\begin{eqnarray}
t_{S_2}: A_1\ &\rightarrow& -2p_1\cdot k_2\, ,
\nonumber\\
t_{S_2}: A_2\ &\rightarrow& A_2\ ,
\nonumber\\
t_{S_2}: A_3\ &\rightarrow& 2p_2\cdot k_2\, ,
\nonumber\\
t_{S_2}: A_4\ &\rightarrow& t\, .
\label{eq:fact_subtract}
\end{eqnarray}
Here, $A_2$, in which every term behaves as $\delta^2$, is kept inact, while only the order $\delta$ terms are kept in
$A_1$ and $A_3$, while the order $\delta^0$ term is kept in $A_4$, all in the $k_2^\mu\sim \delta$ limit, $\mu=0 \dots 3$.
However, in principle, we are allowed to choose subleading terms in  the $\delta$
expansion differently, and for this discussion we will find another choice convenient, in which the {\it only} approximation is to neglect $k_2$ on the off-shell line,
\begin{eqnarray}
&t_{S_2}&:\ A_i\rightarrow A_i\ ,\ i=1,\, 2,\, 3\, ,
\nonumber\\
&t_{S_2}&:\  A_4\ \rightarrow t\, .
\label{eq:fact_subtract-A}
\end{eqnarray}
Clearly, this choice improves the ultraviolet behavior of the resulting expression, by keeping $k_2^2$ terms in three of the denominators \cite{Nagy:2003qn}.   It also results in a better approximation in the collinear regions, as we shall see below.

In a hopefully clear notation, we label the
combination of the original diagram and the particular counterterm defined by Eq.\ (\ref{eq:fact_subtract-A}) 
as Box$_{R1}$ (where
$R1$  simply denotes the remainder after the first subtraction).   Exhibiting the counterterm explicitly, we have
\begin{eqnarray}
\label{eq:Box_R1}
Box_{\rm R1}\ =\ 
(1\, -\, t_{S_2})\, {\rm Box}  &\equiv&
  \int \frac{d^dk_1}{i \pi^{\frac d 2}}
 \frac{1}{A_1 A_2 A_3 A_4}\ -\ \frac{1}{t}\ \int \frac{d^dk_1}{i \pi^{\frac d 2}}
 \frac{1}{A_1 A_2 A_3}
\nonumber\\
&=&  \int \frac{d^dk_1}{i \pi^{\frac d 2}}
\frac{1-\frac{A_4}{t}}{A_1 A_2 A_3 A_4}\, .
\end{eqnarray} 
This subtraction is certainly one of the possible choices that guarantees
that the integral is free of the soft singularity as $k_2 \to 0$.  
The  counterterm in Eq.~(\ref{eq:Box_R1}) is chosen 
according to the prescription of 
Ref.~\cite{Nagy:2003qn}, in which the
denominators of eikonal propagators are not linearized.  The
advantages of the prescription of Ref.~\cite{Nagy:2003qn} are, first,
that soft counterterms do not
introduce spurious UV divergences and second, that they can be 
integrated analytically with standard methods.    

The integrand of Eq~(\ref{eq:Box_R1}) is still divergent in other
regions of the integration domain as, for example, in the remaining 
$k_i^\mu \sim \delta \to 0, \ i=1,3,4$ soft limits.  We subtract
these additional soft singularities sequentially, in the same manner
as above. This process is particularly simple because each of the three
remaining soft limits requires the denominator $A_4$ to vanish for
a divergent contribution in four dimensions.      Indeed, none of
the soft subtraction terms have further soft singularities, and all remaining 
soft divergences are in the first term in Eq.\ (\ref{eq:Box_R1}).

The resulting integral, subtracted for each of its four soft singularities thus 
has four separate subtractions, and takes the form,
\begin{equation}
\label{eq:Box_R}
{\rm Box}_{R}  \equiv \left(1\, -\, \sum_{i=1}^4 t_{S_i}\right) {\rm Box}\ = \int \frac{d^dk_1}{i \pi^{\frac d 2}}
\frac{
N_{\rm Box}
}{A_1 A_2 A_3 A_4},  
\end{equation} 
with 
\begin{equation}
N_{\rm Box}  = 1-\frac{A_{24}}{t} -\frac{A_{13}}{s}   \, .
\end{equation}
It is easy to verify that this integral is not singular at any of the $k_i^\mu \to 0$ soft limits.  

The subtraction in (\ref{eq:Box_R1}) associated with the
$k_2=0$ singularity, for example, is simply $1/t$ times a scalar triangle.   When
regulated dimensionally, the explicit expression for the subtraction is easily integrated, and the four
subtraction terms give
\begin{eqnarray}
\label{eq:soft-sub-result}
t_{S_2}\; {\rm Box}(s,t,\epsilon)\ = t_{S_4}\; {\rm Box}(s,t,\epsilon)\ &=&\ \frac{c_\Gamma}{st \epsilon^2}
(-s)^{-\epsilon}
\nonumber\\[2mm]
t_{S_1}\; {\rm Box}(s,t,\epsilon)\ = t_{S_3}\; {\rm Box}(s,t,\epsilon)\ &=&\ \frac{c_\Gamma}{st \epsilon^2}
  (-t)^{-\epsilon}\, . 
\end{eqnarray}
In Eq.\ (\ref{eq:Box_R}), these terms reproduce and cancel
all double and single poles in the one-loop box, as given in Eq.\ (\ref{eq:box-function}).   Evidently, the
soft subtractions defined as above reproduce all the collinear as well as the soft singularities for 
the particular case of the scalar box.

Turning our attention to the collinear singular limits, as for
example in  Eq.\ (\ref{eq:a-collinear-limit}), we easily confirm that no further subtractions are necessary.
The straightforward application of our method, however, would remove a remaining collinear singularity term by term, by adding an additional subtraction, determined by the collinear behavior of the soft-subtracted integral (\ref{eq:Box_R}).
As noted above, there is some freedom in choosing the subtraction, or counterterm, as long as it matches the singular behavior of the sum of terms in Eq.\ (\ref{eq:Box_R}), and produces no new leading pinch surface. 

Consider the limit in which the loop momentum becomes collinear to external momentum $p_1$.
For this example,  we illustrate one of the forms of collinear counterterms that we shall use below.   The subtraction
acts by keeping the leading finite ($\delta^0$) term in the (two) denominators that are off-shell in this collinear region
($A_3$ and $A_4$), and
the full momentum dependence of the on-shell, collinear denominators ($A_1$ and $A_2$), along with the leading behavior of each term in the numerator $N_{\rm Box}$, Eq.\ (\ref{eq:Box_R}),
that defines the sum of soft subtractions, evaluated at the pinch surface, $k_1=xp_1$, $0<x<1$.   
Representing the action of the $p_1$-collinear approximation by $t_{C_1}$, we have, in particular,
\begin{eqnarray}
t_{C_1}\, A_1\ &=&\ A_1\, ,
\nonumber\\[2mm]
 t_{C_1}\, A_2\ &=&\ A_2\, ,
\nonumber\\[2mm]
 t_{C_1}\, A_3\ &=&\ (1-x) s\, ,
 \nonumber\\[2mm]
 t_{C_1}\, A_4\ &=&\ x t\, .
\end{eqnarray}
 When acting on each of the terms of $N_{\rm Box}$, however, the resulting integral, which has only two full
 denominators, is ultraviolet divergent.    Here, we shall avoid introducing such induced divergences by
 adopting a slight variant of the collinear subtraction introduced in Ref.\ \cite{Nagy:2003qn}.
To be specific, we can introduce an extra factor that approaches unity
at the relevant pinch surface, but which regulates ultraviolet behavior.  
\begin{eqnarray}
\label{eq:tCO-Box}
t_{C_1}\, {\rm Box}  &\equiv&\ 
\int \frac{d^dk_1}{i \pi^{\frac d 2}}\
\left( \frac{1}{A_1}\ -\ \frac{1}{A_1-\mu^2} \right)\; \frac{1}{A_2}\; 
\left[
\frac{1}{s t x_1 (1-x_1)}
\right]
\nonumber \\[2mm]
&=&\ 
\int \frac{d^dk_1}{i \pi^{\frac d 2}}
\left[
\frac{ \frac{\mu^2}{\mu^2-A_1}}{A_1 A_2 s t x_1 (1-x_1)}
\right]\, .
\end{eqnarray}
In the nested approach, we apply the same collinear subtractions to the soft subtraction terms.\footnote{Compared to Ref.\ \cite{Nagy:2003qn}, we do not symmetrize
in the two collinear denominators for each region.  This is  a convention, and will not affect the nature of the results below. }
Treating the remaining collinear regions in the same fashion, the full subtraction is
\begin{eqnarray}
\label{eq:Box_Rprime}
(1- \sum_{i=1}^4 t_{CO\, i} )\ {\rm Box}_{R^\prime}  &\equiv& \int \frac{d^dk_1}{i \pi^{\frac d 2}}
\left[ \frac{ N_{\rm Box}}{A_1 A_2 A_3 A_4}\ -\ \frac{ \frac{\mu^2}{\mu^2-A_1}
{N_{\rm Box} \big |_{k_1 =- x_1 p_1}}}{A_1 A_2 s t x_1 (1-x_1)}
\right].   
\end{eqnarray} 
We expect, of course, that since the soft subtractions 
already cancel all singularities, any term-by-term collinear singularities must likewise cancel among themselves.
This is indeed the case, because non-zero terms in $N_{\rm Box}$ cancel in the
collinear limit for $p_1$ (where $A_1=A_2=0)$, 
\begin{eqnarray}
\label{eq:N-box}
\left. N_{\rm Box} \right|_{k_1 =- x_1 p_1}\ &=&\  \left[ 1 - \frac{A_{13}}{s} - \frac{A_{24}}{t} \right ]\big |_{k_1 =- x_1 p_1}
\nonumber\\
&=& 1 - (1-x_1)-x_1
\nonumber\\
&=&\ 0\, .
\end{eqnarray}
A similar cancellation holds for the remaining three collinear limits.
Thus, for the particular case of the 
one-loop box, we need no
further subtractions for collinear singularities, once we have
introduced counterterms for the soft singular limits as in Eq.\ (\ref{eq:Box_R}). 

We have thus constructed an integral,
Eq.~\eqref{eq:Box_R} that is free of all soft and collinear
singularities. At this stage, we can set the dimension to $d=4$
exactly and perform the loop integral numerically.  
It is important to note that the integral of Eq.~\eqref{eq:Box_R} has further non-pinched
singularities.   Examples are configurations that involve
elastic scattering, if, for example, external particles with momenta $p_1$ and $p_2$ exchange
a non-zero spacelike momentum on line $k_2$ to scatter into an intermediate state with $k_1^2=k_3^2=0$.
  Such singularities, however, can be avoided by appropriate contour
deformation techniques, as suggested for example in
Refs.~\cite{Nagy:2006xy,Gong:2008ww,Becker:2012nf,Becker:2012aqa,Becker:2012bi,Bierenbaum:2010cy,Buchta:2015wna}. 

Although in general we would expect to evaluate the remainder with
nunerical methods, as an illustration in the one-loop case  
of Eq.~(\ref{eq:Box_R}), we  can 
introduce Feynman parameters, ``complete the square''  in the
loop-momentum and drop numerator terms in odd powers of 
 the loop-momentum, which integrate trivially to zero.  We find,
\begin{equation}
{\rm Box}_{R}  = -2 \frac{s+t}{s t} \Gamma(4)\int \frac{d^4k}{i \pi^{2}}
d x_1 dx_2 d x_3 dx_4 \delta(1- x_{1234})
\frac{
k^2 - \Delta
}{
\left[ k^2 + \Delta + i 0\right]^4
} 
\end{equation}
where 
\begin{equation}
\Delta = x_1 x_3 s +  x_2 x_4 t.
\end{equation}
We integrate out the loop-momentum, resulting in
\begin{equation}
{\rm Box}_{R}  = -2 \frac{s+t}{st} \int 
d x_1 dx_2 d x_3 dx_4 \delta(1- x_{1234})
\frac{1}{ \Delta + i 0}\, .
\end{equation}
The integration can be performed by standard methods, yielding the finite result: 
\begin{equation}
{\rm Box}_{R}  =  -\frac{1}{st}\left[ \pi^2 + \ln^2\left( \frac t
    s \right)\right]. 
\end{equation}
This is indeed the correct contribution to the finite part of the
integral, as is found by comparing Eqs.~(\ref{eq:box-function}) and (\ref{eq:soft-sub-result}). 

In summary, for our introductory one-loop example,
the method of nested subtractions employed here 
yields the same separation of finite and divergent terms as the method of  Ref.~\cite{Nagy:2003qn}. 
In the following, we will demonstrate that nested subtractions also allow us to treat 
infrared divergences in non-trivial, two-loop examples.  


\section{Application to two-loop scalar integrals}
\label{sec:examples}

As noted above, it has been shown \cite{Collins:2011zzd,Erdogan:2014gha}  that  we can remove the infrared singularities of 
 multi-loop integrals for hard scattering processes
with suitable nested subtractions, which we have described in the previous section.  
However, beyond one-loop in multi-leg amplitudes,  
we are not aware of
a practical construction that realizes this potential 
in the literature.  
In this section, we  apply for the first time our method of nested
subtractions at two loops. 

We will focus on  two-loop integrals with
four external legs which, for light-like external momenta, already have a 
complicated singular structure. Explicitly, we will test that we can
render integrable in $d=4$ dimensions  the ``diagonal-box'',  
the ``bubble-box'', the ``planar double-box'' and the  ``crossed
double-box''.  These integrals represent Feynman diagrams for
the scattering of massless scalar particles. In addition, they are the most 
complicated {\it master integrals}
which appear in all $2 \to 2$ scattering processes in massless QCD.  
We believe that the set of integrals that we examine here  serves
two purposes:  giving a pedagogical introduction to our technique at
two loops, and testing it thoroughly in non-trivial applications.
In particular, the  planar and crossed double-box integrals 
have poles in the dimensional
regulator of the maximum power, $1/\epsilon^4$, as they 
possess all the infrared singularities that are anticipated at
two loops.  Largely due to their complicated singular structure,  the analytic
evaluation of the planar and crossed  double-box integrals was not
amenable to traditional techniques, and was only
achieved for the first time when  Smirnov~\cite{Smirnov:1999gc} and
Tausk~\cite{Tausk:1999vh} developed powerful Mellin-Barnes methods.
 
In this section, we will show that the analytic structure of the
$1/\epsilon$ poles of our two-loop examples
can be derived in a simple way, by integrating less complicated
counterterm integrals.    In addition, our counterterm 
subtractions will render the remainders of the integrands free of any
local singularities, and therefore amenable to direct integration
methods in exactly $d=4$ dimensions.  

We begin our discussion with two relatively simple cases, the ``diagonal" and ``bubble" boxes.
In these cases, no more than two nested subtractions are necessary, and the pattern follows
the general considerations outlined in the previous section.   We will use these examples, however,
to illustrate convenient choices of finite parts for collinear subtractions.   
We then turn to the more complex cases of the planar and nonplanar double boxes.



\subsection{Subtraction for the diagonal-box} 
\label{sec:diagbox-massless}
\begin{figure}[h]
\begin{center}
\includegraphics[width=0.5\textwidth]{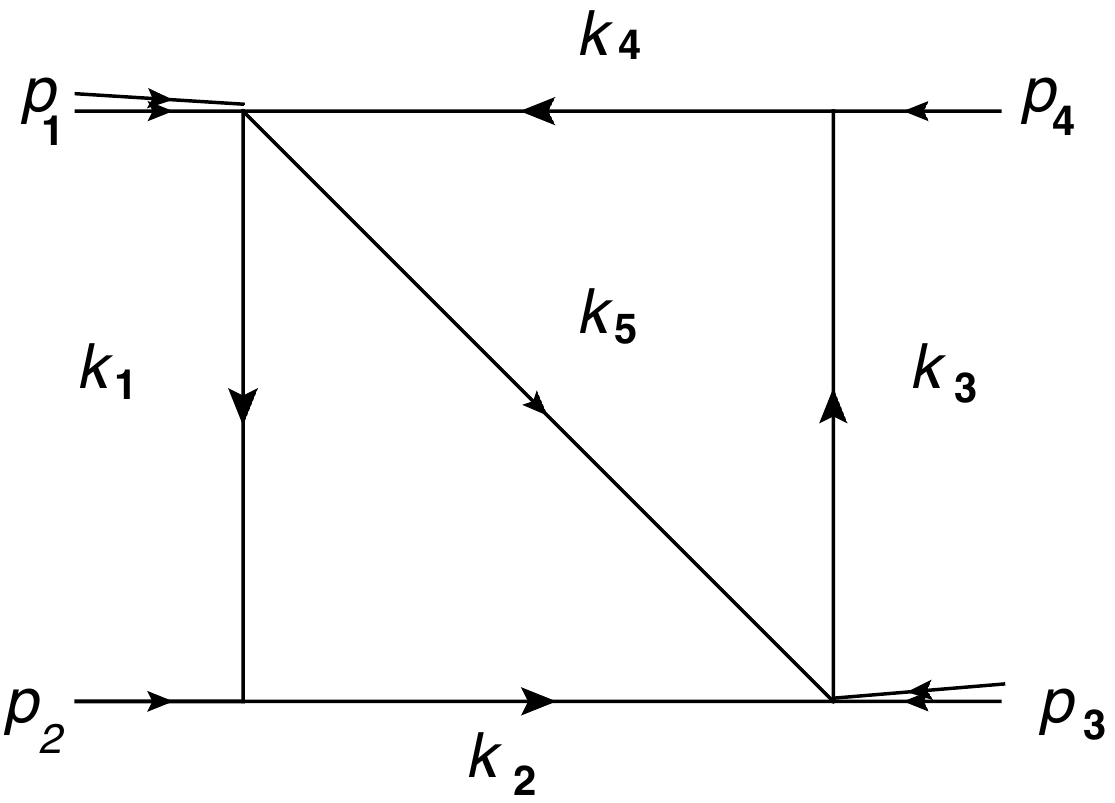}
\caption{\label{fig:diagonal-box} The two-loop diagonal-box}
\end{center}
\end{figure}
As our first two-loop application, we choose an example with only collinear singularities.
Consider the diagonal-box integral, defined as
\begin{equation}
\label{eq:diagbox_definition}
\Dbox \equiv \int \frac{d^dk_1}{i \pi^\dhalf}
\frac{d^dk_4}{i\pi^\dhalf}  \frac{1}{A_1 A_2 A_3 A_4  A_5}\, , \quad 
\end{equation}
with 
\begin{equation}
A_i = k_i^2 + i0\, . 
\end{equation}
The momenta $k_i$ of the propagators are depicted in
Fig.~\ref{fig:diagonal-box}. One can concretely identify loop momenta with the lines $k_1$ and $k_4$, so that
\begin{equation}
\label{eq:routing_diagb}
k_1=l, \ \ k_2=l+p_{2},\  \ k_3=k+p_{123}, \ \ k_4=k,\ \  k_5=k-l+p_1. 
\end{equation}
The kinematics of the external momenta $p_i$ are,
\begin{equation}
\sum_{i=1}^4 p_i =0,  \quad p_2^2=p_4^2=0, p_1^2=m_1^2, p_3^2=m_3^2, \quad p_{12}^2=p_{34}^2=s, \quad
p_{23}^2=p_{14}^2=t. 
\end{equation}
We have taken the $p_1,p_3$ momenta to be off-shell.   Our study
carries through unchanged, however, in the case that one or both of them go
on-shell. 
 
By inspecting all 
pinch surfaces, using the power counting of Eq.\ (\ref{eq:power-ctg-scalar}), we find that the diagonal-box has only collinear
singularities, which we sort according to
increasing volumes of their regions:  
\begin{itemize}
\item {\it Two-collinear pairs:} $C_{1 || 2} C_{4 ||4}$, in which the
  internal momenta $k_1,k_4$ are simultaneously parallel to the external momenta
  $p_2 ,p_4$ respectively.   
  In the notation of Eq.\ (\ref{eq:collinear_momentum}), we parameterize the loop momenta  as
\begin{equation}
k_1 = - x_2 p_2 + \beta_2 \eta_2 + k_{\perp_2}, 
k_4 =  x_4 p_4 + \beta_4 \eta_4 + k_{\perp_4},  
\end{equation}
in this region, where 
\begin{equation}
x_2,x_4 \sim {\cal O}(1), \quad  \beta_{2}, \beta_4 \sim {\cal O}(\delta),
k_{\perp_2}, k_{\perp_4} \sim {\cal O}\left(\delta^{\frac 1 2}\right),
\end{equation}
and $k_{\perp_{i}} \cdot p_i=k_{\perp_{i}} \cdot \eta_i=0, \; p_i
\cdot \eta_i \neq 0$. 
\item {\it Single-Collinear:} 
\begin{itemize}
\item $C_{1||2}$, in which the internal
  momentum $k_1$ is parallel to the external momentum $p_2$, 
\begin{equation}
k_1 = - x_2 p_2 + \beta_2 \eta_2 + k_{\perp_2}, 
\end{equation}
where 
\begin{equation}
x_2 \sim {\cal O}(1), \quad  \beta_{2},  \sim {\cal O}(\delta),
k_{\perp_2}, \sim {\cal O}\left(\delta^{\frac 1 2}\right), k_4 \sim
{\cal O}(1).
\end{equation}
\item $C_{4||4}$, in which the internal
  momentum $k_4$ is parallel to the external momentum $p_4$,
\begin{equation}
k_4 =  x_4 p_4 + \beta_4 \eta_4 + k_{\perp_4},  
\end{equation}
where 
\begin{equation}
x_4 \sim {\cal O}(1), \quad  \beta_4 \sim {\cal O}(\delta),
k_{\perp_4} \sim {\cal O}\left(\delta^{\frac 1 2}\right), 
k_1 \sim {\cal O}(1).
\end{equation}
\end{itemize}
\end{itemize}

In the $C_{1 || 2} C_{4 ||4}$ limit, the momenta 
corresponding to $A_1, A_2, A_3, A_4 \sim {\cal O}(\delta)$ become
on-shell, while $A_5 \to (x_2 p_2 +x_4 p_4 +p_1)^2 \sim {\cal O}(1)$ is
off-shell. We can remove this singularity by introducing a counterterm
that subtracts an approximation to the integrand in this singular
limit.  In the notation of the previous section, we denote this as
\begin{eqnarray}
\label{eq:dbox-R1}
\left. \Dbox\right|_{{\rm R}_1} && =\
\left(1\ -\ t_{C_2C_4}\right) \, D_{\rm box}
\nonumber\\[2mm]
&& \equiv\
\int \frac{d^dk_1}{i \pi^\dhalf}
\frac{d^dk_4}{i\pi^\dhalf}  \Bigg\{ 
\frac{1}{A_1 A_2 A_3 A_4  A_5} 
- 
\frac{1}{A_1 A_2  A_3 A_4} 
\left[ 
\frac{1}{A_5}
\right]_{\begin{array}{l}
{}_{k_4= x_4 p_4,}\\
{}^{ k_1=-x_2 p_2}
\end{array}
}
\Bigg\}\, ,
\nonumber \\ &&
\end{eqnarray}
where $t_{C_2C_4}$ represents the ``two-collinear pairs" approximation shown in the second equality.
Here and below, we employ a notation in which the function inside square brackets is
evaluated at the values of momenta shown in the subscript.   We note that a factor of $\theta(x_i)\theta(1-x_i)$,
which ensures that along the collinear surface the momentum of a lightlike external line is shared by two internal lines, moving in the same direction,
will emerge after loop integrals at these collinear limits.  For now, however, the $x_i$ may be
taken as unconstrained.   In the example at hand, we then have
\bea
\left[ 
\frac{1}{A_5}
\right]_{\begin{array}{l}
{}_{k_4= x_4 p_4,}\\
{}^{ k_1=-x_2 p_2}
\end{array}}
\ &=&\ 
\frac{1} {(x_2 p_2 +x_4 p_4 +p_1)^2}
\nonumber\\[2mm]
&\equiv& \frac{1} {A(x_2,x_4)}\, ,
\eea
where in the second equality, we identify a function that we will encounter
in explicit integrals below,
\begin{equation}
\label{eq:A-def}
A(x, y)= \left(p_1+x p_2+y p_4 \right)^2\, .
\end{equation}
The subtraction in Eq.\ (\ref{eq:dbox-R1}) removes the double collinear limit,
and we take this as our starting point for the removal of the remaining singularities
associated with single-collinear limits.   
We will then turn to the treatment of induced
ultraviolet poles.

We now proceed to remove the divergence due to the $C_{1||2}$ single-collinear 
limit in $\left. \Dbox\right|_{{\rm R}_1}$. In this limit, $A_1, A_2 \sim
{\cal O}(\delta)$ while $A_3, A_4, A_5 \sim {\cal O}(1)$.    To treat this region, we
introduce an additional counterterm to subtract the behavior of the full integrand of $\left. \Dbox\right|_{{\rm R}_1}$
in Eq.\ (\ref{eq:dbox-R1}).   The resulting subtraction automatically avoids over-counting the double-collinear
limit, which is already removed.   In the same notation as Eq.\ (\ref{eq:dbox-R1}), we represent the resulting expression as
\begin{eqnarray}
\left. \Dbox\right|_{{\rm R}_2} && =\
\left( 1\ -\ t_{C_2C_4}\ - t_{C_2}\  +\ t_{C_2}t_{C_2C_4}\  \right) \, D_{\rm box}
\nonumber\\[2mm]
&& \equiv\ \int \frac{d^dk_1}{i \pi^\dhalf}
\frac{d^dk_4}{i\pi^\dhalf}  
\Bigg[
\Bigg\{ 
\frac{1}{A_1 A_2 A_3 A_4  A_5} 
- 
\frac{1}{A_1 A_2  A_3 A_4} 
\left[ s
\frac{1}{A_5}
\right]_{\begin{array}{l}
{}_{k_4= x_4 p_4,}\\
{}^{ k_1=-x_2 p_2}
\end{array}
}
\Bigg\}
\nonumber \\ &&
- \Bigg\{ 
\frac{1}{A_1 A_2}
\left[ 
\frac{1}{A_3 A_4 A_5}
\right]_{k_1=-x_2 p_2}
- 
\frac{1}{A_1 A_2} 
\left[ 
\frac{1}{A_3 A_4}
\right]_{k_1 =-x_2 p_2}
\left[ 
\frac{1}{A_5}
\right]_{\begin{array}{l}
{}_{k_4= x_4 p_4,}\\
{}^{ k_1=-x_2 p_2}
\end{array}
}
\Bigg\} \Bigg]
\nonumber \\ 
&=& 
\int \frac{d^dk_1}{i \pi^\dhalf}
\frac{d^dk_4}{i\pi^\dhalf}  
\Bigg\{ 
\frac{1}{A_1 A_2 A_3 A_4  A_5} 
- 
\frac{1}{A_1 A_2}
\left[ 
\frac{1}{A_3 A_4 A_5}
\right]_{k_1=-x_2 p_2}
\Bigg\}\, .
\end{eqnarray}
In the second equality, we have noted that
 the second and fourth terms on the right of the first equality cancel,  because
 momentum $k_1$  (which becomes parallel to $p_2$ in this limit) flows only through
propagator $A_5$, according to our choice of routing of the momenta in
Eq.~(\ref{eq:routing_diagb}),  so that propagators $A_3,A_4$ are
independent of $k_1$.   In the subtraction notation, this amounts to $t_{C_2}t_{C_2C_4}\ D\ = t_{C_2C_4}\ D.$
We retain denominators $A_3,\, A_4$ inside the square brackets,  as a reminder that they appear with denominator $A_5$ 
in a one-loop integral, evaluated at fixed $k_1=-x_2p_2$.

We remove the remaining single-collinear singularity
$C_{4 ||4}$ from $\left. \Dbox\right|_{{\rm R}_2}$ similarly, giving
 giving
\begin{eqnarray}
\label{eq:Dbox-R3}
\left. \Dbox\right|_{{\rm R}_3} && =\
\left( 1\ -\ t_{C_2C_4}\ - t_{C_2}\ -\ t_{C_4} +\ t_{C_2}t_{C_2C_4}\ +\ t_{C_4}t_{C_2C_4}\right) \, D_{\rm box}
\nonumber\\[2mm]
&&
\int \frac{d^dk_1}{i \pi^\dhalf}
\frac{d^dk_4}{i\pi^\dhalf}  \Bigg\{ 
\frac{1}{A_1 A_2 A_3 A_4  A_5} 
- 
\frac{1}{A_1 A_2} 
\left[ 
\frac{1}{A_3 A_4 A_5}
\right]_{k_1=-x_2 p_2}
\nonumber \\ 
&& 
- 
\frac{1}{A_3 A_4} 
\left[ 
\frac{1}{A_1 A_2 A_5}
\right]_{k_4= x_4 p_4} 
+ 
\frac{1}{A_1 A_2  A_3 A_4} 
\left[ 
\frac{1}{A_5}
\right]_{\begin{array}{l}
{}_{k_4= x_4 p_4,}\\
{}^{ k_1=-x_2 p_2}
\end{array}
}
\Bigg\}\, ,
\nonumber \\ &&
\end{eqnarray}
where again, we use $t_{C_4}t_{C_2C_4}=t_{C_2C_4}$.
At this stage, we have an integral that is free of all infrared
singularities.
 However, the counterterms that we have introduced are
divergent in the ultraviolet limit, just as they were for the
one-loop box treated in the previous section.  For an analysis carried out
purely in dimensional regularization, this would not be a problem, but
since, as above, our goal is to derive integrals that can be evaluated numerically,
we need them to converge in four dimensions.

As an additional step, therefore, we modify our counterterms, so that they depend on an artificial mass $\mu$
in a manner that tames
the ultraviolet behavior of the integrand.   These are a variant of the 
subtraction in Eq.\ (\ref{eq:tCO-Box}) above, still in the spirit of Ref.\ \cite{Nagy:2003qn}.
This gives our final expression for a fully-subtracted diagonal box, now finite in four dimensions.
The subtractions are in the same pattern as in (\ref{eq:Dbox-R3}), but all integrals are now UV convergent,
\begin{eqnarray}
\label{eq:diagb_fin}
\left. \Dbox\right|_{\rm fin} && =
\int \frac{d^dk_1}{i \pi^\dhalf}
\frac{d^dk_4}{i\pi^\dhalf}  \Bigg\{ 
\frac{1}{A_1 A_2 A_3 A_4  A_5} 
 \nonumber \\ 
&& 
- 
\left[ 
\frac{1}{A_1 A_2} - \frac{1}{\left(A_1 - \mu^2 \right) \left(A_2-\mu^2\right)}
\right]
\left[ 
\frac{1}{A_3 A_4 A_5}
\right]_{k_1=-x_2 p_2}
\nonumber \\ 
&& 
- 
\left[ 
\frac{1}{A_3 A_4} - \frac{1}{\left(A_4 - \mu^2 \right) \left(A_3-\mu^2\right)}
\right]
\left[ 
\frac{1}{A_1 A_2 A_5}
\right]_{k_4= x_4 p_4} 
 \nonumber \\ && 
\hspace{-2cm}
+ 
\left[ 
\frac{1}{A_1 A_2 } - \frac{1}{
\left(A_1 - \mu^2 \right) \left(A_2-\mu^2\right) }
\right]
\left[ 
\frac{1}{A_5}
\right]_{\begin{array}{l}
{}_{k_4= x_4 p_4,}\\
{}^{ k_1=-x_2 p_2}
\end{array}}
\left[ 
\frac{1}{A_3 A_4} - \frac{1}{
 \left(A_3 - \mu^2 \right) \left(A_4-\mu^2\right)}
\right]
\Bigg\}\, .
\nonumber \\ &&
\end{eqnarray}
In this expression, we have added to each subtraction term
in (\ref{eq:Dbox-R3}) an IR finite adjustment, in which mass dependence
is introduced in the denominators that become collinear.   Of
course, this introduces poles associated with the new
denominators.    As long as $\mu$ is finite, however, these poles
produce no new pinches,  because lines $p_2$ and $p_4$ are
lightlike.    The full diagonal box equals $\left. \Dbox\right|_{\rm fin}$ plus the 
 counterterms in (\ref{eq:diagb_fin}).
The integral $\left. \Dbox\right|_{\rm fin}$ of Eq.~(\ref{eq:diagb_fin})
can now be evaluated numerically (at least in principle) or
analytically  in exactly d=4 dimensions, since it is free of all
divergences.   

In fact, for this case, 
evaluation in dimensional regularization is particularly simple, because
in dimensional regularization all mass-independent counterterms 
include scaleless integrals that vanish.   In this way, of the nine terms in 
Eq.\ (\ref{eq:diagb_fin}), only the first, third,
fifth and the final term with four massive denominators survive.
Of course, by using dimensional regularization we abandon 
the use of point-by-point cancellation in Eq.\ (\ref{eq:diagb_fin}).
Nevertheless, it will enable us to confirm the finiteness of the
full subtracted form.   These integrals will also come in handy in our discussion 
of mass-dependent integrals in Sec.\ \ref{sec:smallmass}.

We thus proceed to evaluate both the finite part and the singular
parts of the original two-loop integral.  The latter emerge from the
integration of the counterterms, which we will perform
in non-integer $d=4-2\epsilon$ dimensions.  Not surprisingly, the
integrations for the counterterms are simpler than the
integration of the original integral.
\begin{figure}
\begin{center}
\includegraphics[width=0.5\textwidth]{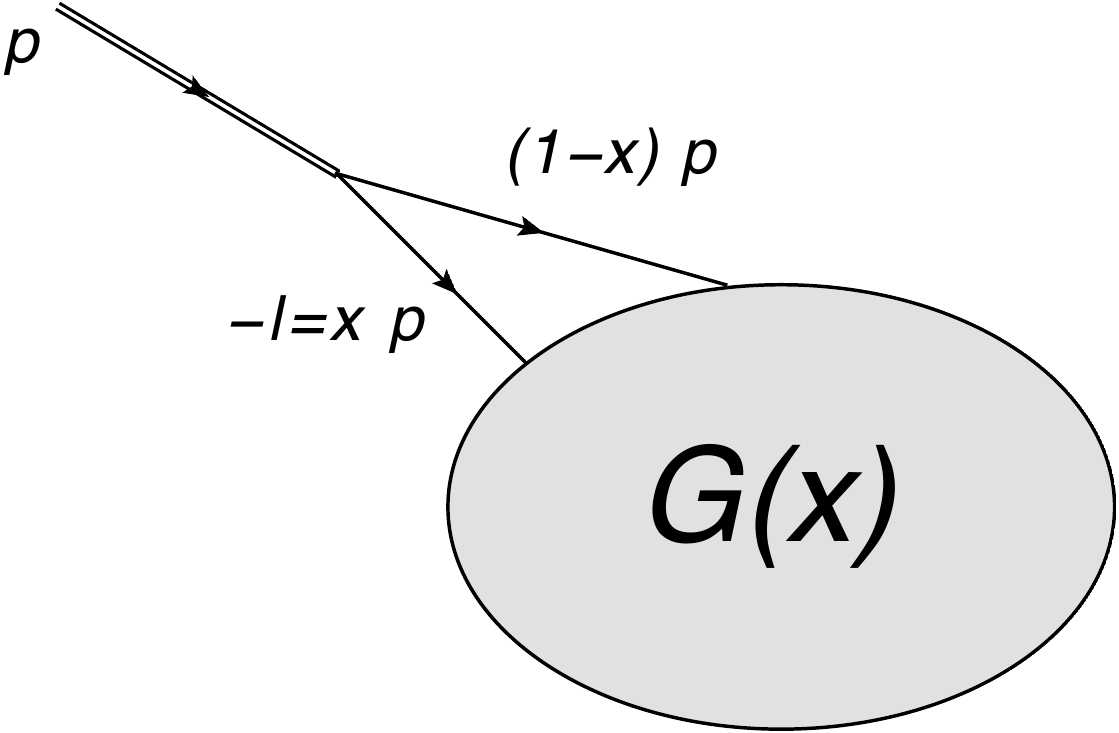}
\end{center}
\caption{
\label{fig:gencollint}
A generic representation of a collinear configuration.}
\end{figure}  

The collinear counterterms that UV-regulate the diagonal box
require  integrals of the generic form  
\begin{equation}
\label{eq:gencollint}
I_G(m,M) \equiv \int \frac{d^d l}{i \pi^{\frac d 2}} \frac{G(x(l))}{\left[
    l^2-m^2\right]  
\left[  (l+p)^2 - M^2 \right]} \, ,
\end{equation}
where $p$ is an on-shell external momentum, $l$ and $l+p$ are the momenta
of the propagators attached to $p$ and $x(l) = -\frac{l \cdot \eta_p}{p
  \cdot \eta_p}$ is the fraction of $p$  that is carried by $l$. $G(x(l))$
is the internal subgraph that is attached to the propagators $l, l+p$, as illustrated
in Fig.~\ref{fig:gencollint}.

The masses $m,M$ in Eq.\ (\ref{eq:gencollint}) take non-zero values only  in
the terms of Eq.\ (\ref{eq:diagb_fin}) that have been added to
regulate the ultraviolet limit of the collinear counterterms, in which case, $m=M=\mu$.
We can treat all these integrals in the same fashion, by first
introducing a subintegral that is differential in the momentum fraction
$x$,  
\begin{equation}
\label{eq:gencollint_x}
I_c(x, m, M) \equiv \int \frac{d^d l}{i  \pi^{\frac d 2}} 
\frac{\delta\left( x + \frac{2 l \cdot \eta}{2 p \cdot \eta}\right)}{\left[l^2-m^2 \right]
  \left[ (l+p)^2-M^2 \right]},  
\end{equation}
in terms of which,
\begin{equation}
\label{eq:collconv}
I_G(m,M) = \int_{-\infty}^\infty dx\, I_c[x, m, M] G(x)\, . 
\end{equation}
The  integral of Eq.~(\ref{eq:gencollint_x}) at fixed $x$ can be computed
easily by using light-cone integration variables and Cauchy's
theorem. It reads 
\begin{equation}
\label{eq:gencollint_x_result}
I_c[x, m, M] = \frac{\Gamma(1+\epsilon)}{\epsilon}\left[ x m^2 +(1-x)
  M^2 \right]^{-\epsilon} \Theta\left( 0  \leq x \leq 1\right). 
\end{equation}
Therefore, we find that 
\begin{equation}
\label{eq:collconv_result}
I_G = \frac{\Gamma(1+\epsilon)}{\epsilon}
\int_{0}^1 dx \left[ x m^2 +(1-x) M^2 \right]^{-\epsilon}  G(x). 
\end{equation}
Here, the explicit denominator reduces to a constant whenever the masses $m$ and $M$ are equal (to $\mu$ in the case at hand).

For example, 
Eq.\ (\ref{eq:collconv_result}) allows us to evaluate
the double-collinear counterterm, for which
$G=[1/A_5]$, for both the $k_1$ and $k_4$ integrals, which result in $x_1\equiv x$ and $x_4\equiv y$ integrations, respectively.  For this term,
we find in this way,  
\begin{eqnarray}
\label{eq:k1k4-int}
&& 
\int \frac{d^dk_1}{i \pi^\dhalf}
\frac{d^dk_4}{i\pi^\dhalf}  \frac{1}{
\left(A_1 - \mu^2 \right) \left(A_2-\mu^2\right) \left(A_4 - \mu^2 \right) \left(A_3-\mu^2\right)}
\left[ 
\frac{1}{A_5}
\right]_{\begin{array}{l}
{}_{k_4= x_4 p_4,}\\
{}^{ k_1=-x_2 p_2}
\end{array}
}
\nonumber \\  
&& 
= \frac{\Gamma(1+\epsilon)^2}{\epsilon^2} (\mu^2)^{-2 \epsilon}
\int_0^1 dx dy \frac{1}{A(x,y)}\, ,
\end{eqnarray}
where $A(x,y)$ is given by Eq.\ (\ref{eq:A-def}).   
The remaining integrals 
can then be done in a straightforward manner analytically in terms of rank-two 
polylogarithmic functions.  
We define 
the kinematic variables,
\begin{eqnarray}
 u =m_1^2+m_3^2-s-t, 
&& K = m_1^2 m_3^2-st \, ,  \nonumber  \\[2mm]
v_1 = \frac{u m_1^2}{K},  \;  v_3 = \frac{u m_3^2}{K},
&&
v_s = \frac{u s }{K}, \;   v_t = \frac{u t}{K}\, ,
\end{eqnarray}  
and the scale $\mu$-dependent logarithmic function,
\begin{equation}
L_\mu(z) \equiv \log \left(-\frac{z}{\mu^2} \right)\, .
\end{equation}
In these terms, we find 
\begin{eqnarray}
\label{eq:dbox_ep2_polylog}
u \int_0^1 dx dy \frac{1}{A(x,y)}  
&=&  -{\rm Li}_2(v_1)-{\rm Li}_2(v_3) + {\rm Li}_2(v_s) + {\rm Li}_2(v_t)
- \ln(1-v_1)L_\mu(m_1^2)
\nonumber \\ && \hspace{-3cm}
- \ln(1-v_3)L_\mu(m_3^2) 
+ \ln(1-v_s) L_\mu(s)
+ \ln(1-v_t) L_\mu(t)\, ,
\end{eqnarray}
which will appear as the coefficient of the double-pole counterterm for the diagonal box.

The integration of the single-collinear counterterms
requires one more integral, which arises from the off-shell triangle
that is opposite to the counterterm.   
Consider, for example, the triangle integral corresponding to loop momentum $l=k_1$ in
Fig.\ \ref{fig:diagonal-box}, evaluated at $k_4=x_4p_4$, as it appears in the third subtraction term in Eq.\ (\ref{eq:diagb_fin}).
The relevant integral now has three propagators,
\bea
I_\Delta(p_1,x_4p_4)\ &=&\ \int \frac{d^dl}{i\pi^{\frac{d}{2}}}\, \frac{1}{l^2+i0} \frac{1}{(p_2+l)^2+i0}\, \frac{1}{(p_1+x_4p_4 - l)^2+i0}
\nonumber\\[2mm]
&=&\ B(1-\epsilon,-\epsilon)\, \int_0^1\ \frac{dy}{[A(y,x_4)]^{1+\epsilon}}\, ,
\label{eq:Delta-def}
\eea
which is of the same form as the two-propagator integral in Eq.\ (\ref{eq:gencollint_x})
giving Eq.\ (\ref{eq:k1k4-int}), but now with an $\epsilon$-dependent 
factor of $A(y,x_4)$ instead of the factor $\mu^{-2\epsilon}$ times $A$.   
The result in Eq.\ (\ref{eq:Delta-def}) 
is
easily shown by using Feynman parameters, first to combine the denominators of the 
two lines attached to the lightlike external line, and
then combining that result with the remaining denominator, for line $k_5$ in the figure 
\cite{Anastasiou:2011zk}. The same method applies to the original diagram, because doing either
one-loop integral as a triangle, results in another integral, which differs only in having one denominator raised to the power $1+\epsilon$.

In the computation of the finite part, Eq.\ (\ref{eq:diagb_fin}), the ``single-collinear" term from (\ref{eq:Delta-def}) contributes a single pole
from the expansion of $A^{-1-\epsilon}$ that
can also be computed easily in closed form,
in terms of rank-three polylogarithmic functions, 
\begin{eqnarray}
\label{eq:dbox_ep_polylog}
u \int_0^1 dx dy \frac{\ln \left(\frac{A(x,y)}{\mu^2}\right)}{A(x,y)}  
&=& 
{\rm Li}_3(v_1)+{\rm Li}_3(v_3)- {\rm Li}_3(v_s) -{\rm Li}_3(v_t)
\nonumber \\ &&  \hspace{-3cm}
-L_\mu(m_1^2) {\rm Li}_2(v_1)
- L_\mu(m_3^2)  {\rm Li}_2(v_3)+L_\mu(s){\rm Li}_2(v_s)
                +L_\mu(t) {\rm Li}_2(v_t)
\nonumber \\ &&  \hspace{-3cm}
+ \frac 1 2  \ln(1-v_1)L_\mu^2(m_1^2)
+ \frac 1 2 \ln(1-v_3)L_\mu^2(m_3^2) 
\nonumber \\ &&  \hspace{-3cm}
- \frac 1 2 \ln(1-v_s) L_\mu^2(s)
- \frac 1 2 \ln(1-v_t) L_\mu^2(t)\, .
\end{eqnarray}
Quite generally, for collinear limits in two-loop diagrams, the kernel $G(x)$ corresponds
to a one-loop or a tree subgraph. $G(x)$ is therefore a rational
function of $x$ for trees and a function of polylogarithms of $x$ for
one-loop subgraphs. The method of nested subtractions leads to
functions $G(x)$, which contain no other subdivergences.  Therefore,
integrands of the form of Eq.~(\ref{eq:collconv_result}) can be expanded as a Taylor
series in $\epsilon$, whose coefficients can be integrated either
analytically or, alternatively, numerically.    

Having discussed the integration of the divergent counterterms, 
we return to the evaluation of the finite remainder of Eq.~(\ref{eq:diagb_fin}),
which in this case can be performed in exactly four dimensions. 
 We envisage that finite remainders of two-loop integrals after the application of
nested subtractions are integrated numerically in momentum
space, after appropriate contour deformations away from  non-pinched
singularities are applied. 
We emphasize again that the development of an efficient numerical method requires further
study, a problem that we will not address here.
A method that achieves this purpose for
generic multi-loop integrals has been presented in 
Ref~\cite{Becker:2012bi}. 


For the full finite part, including the original diagram, we have from the above,
\begin{eqnarray}
\left. \Dbox\right|_{\rm fin} &&
                                 =-\frac{\Gamma(1+\epsilon)^2}{\epsilon^2}
                                 \int_0^1 \frac{dx dy}{A(x,y)}
\left[
\frac{
\Gamma(1+2\epsilon) \Gamma(1-\epsilon)^3
}{\
\Gamma(1+e)^2 \Gamma(1-3\epsilon)}
A(x,y)^{-2 \epsilon}
\right. 
\nonumber \\ 
&& \left.
-2 \frac{\Gamma(1-\epsilon)^2}{\Gamma(1-2\epsilon)^2}
\left( \mu^2  A(x,y) \right)^{-\epsilon}
+\left( \mu^2\right)^{-2\epsilon}                                  
\right] \nonumber \\ 
&=& 
\int_{0}^1 dx dy \frac{\log^2\left(
   -\frac{A(x,y)}{\mu^2}\right)}{A(x,y)} +{\cal O}(\epsilon)\, ,
\end{eqnarray} 
with $A(x,y)$ given in Eq.\ (\ref{eq:A-def}).   The first term
in brackets on the right of the first equality
 is the full diagram, the second term is the
result of single-collinear subtractions, and the third term is from the double-collinear subtraction.
This expression is manifestly finite in $d=4$ dimensions and  can also be easily 
integrated analytically in terms  of logarithms and polylogarithms
(see, for example, Appendix D of Ref.~\cite{Anastasiou:2013srw}). 
The analytic result for the finite remainder of the diagonal-box integral reads 
\begin{eqnarray}
\label{eq:dbox_fin_polylog}
 u  \left. \Dbox\right|_{\rm fin}(\mu) &=& 
2 {\rm Li}_4(v_1)+2 {\rm Li}_4(v_3) - 2  {\rm Li}_4(v_s)- 2 {\rm Li}_4(v_t) 
\nonumber \\ && \hspace{-2.4cm}
- 2 {\rm Li}_3(v_1)L_\mu(m_1^2)- 2{\rm Li}_3(v_3) L_\mu(m_3^2) +2  {\rm
  Li}_3(v_s) L_\mu(s)+2 {\rm Li}_3(v_t) L_\mu(t)
\nonumber \\ && \hspace{-2.4cm}
+ {\rm Li}_2(v_1)L_\mu^2(m_1^2)+ {\rm Li}_2(v_3) L_\mu^2(m_3^2) - {\rm
  Li}_2(v_s) L_\mu^2(s)- {\rm Li}_2(v_t) L_\mu^2(t)
\nonumber \\ && \hspace{-2.4cm}
+ \frac 1 3 \ln(1-v_1)L_\mu^3(m_1^2)+ \frac 1 3 \ln(1-v_3)
                L_\mu^3(m_3^2) - \frac 1 3 \ln(1-v_s) L_\mu^3(s)
\nonumber \\ && \hspace{-2.4cm}
- \frac 1 3 \ln(1-v_t) L_\mu^3(t)\, .
\end{eqnarray}
The limit of $m_1,m_2 \to 0$ can be taken smoothly in Eq.~(\ref{eq:dbox_fin_polylog}). We
have checked that in that limit the above result, when combined with
the integrated counterterms,  agrees with the
analytical results of  Refs.~\cite{Smirnov:1999wz,Anastasiou:1999bn}
for $m_1=m_3=0$. Finally, we would like to comment that the terms
proportional to $\log^3(\mu)$ in Eq.~(\ref{eq:dbox_fin_polylog}), 
$\log^2(\mu)$ in Eq.~(\ref{eq:dbox_ep2_polylog})
and $\log(\mu)$ in Eq.~(\ref{eq:dbox_ep_polylog})
all cancel. This is in accordance with expectations, since the 
strongest singularity is due to two-collinear pairs capable of
producing at most $1/\epsilon^2$ poles and consequently at most
$\log^2(\mu)$ terms in the finite part of the integral.

\subsection{Subtraction for the bubble-box} 
\begin{figure}[h]
\begin{center}
\includegraphics[width=0.5\textwidth]{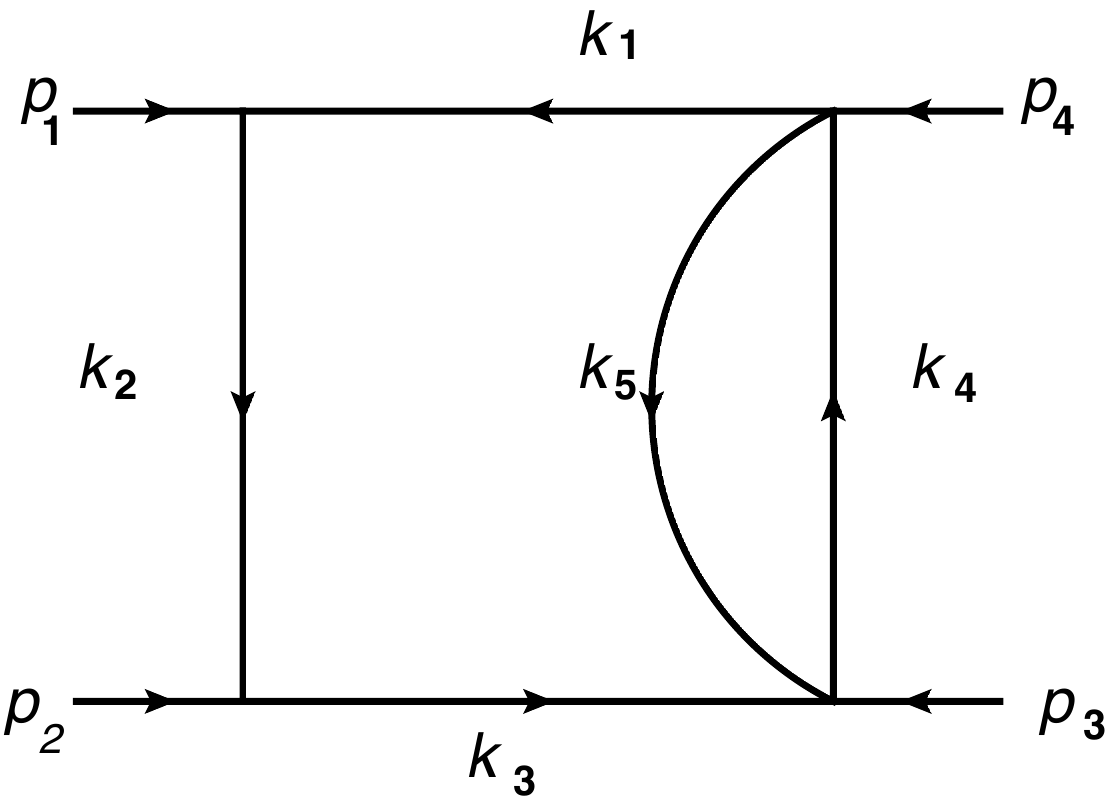}
\caption{\label{fig:bubble-box} The two-loop bubble-box}
\end{center}
\end{figure}
We now consider the bubble-box two-loop integral, which has collinear and soft IR divergences,
in addition to a UV-divergent subdiagram.   We give fewer details in this case, because the necessary reasoning 
is very similar to the one-loop and diagonal box diagrams, and is easily seen to give the desired form.

The bubble-box diagram was computed analytically in  Ref~\cite{Anastasiou:1999bn}. 
It is defined as
\begin{equation}
\label{eq:bubblebox_definition}
\Bbox \equiv \int \frac{d^dk_2}{i \pi^\dhalf}
\frac{d^dk_5}{i\pi^\dhalf}  \frac{1}{A_1 A_2 A_3 A_4  A_5}, \quad 
\end{equation}
with 
\begin{equation}
A_i = k_i^2 + i0. 
\end{equation}
The momenta $k_i$ of the propagators are depicted in
Fig.~\ref{fig:bubble-box}. One can concretely choose
\begin{equation}
k_1=l, k_2=l+p_{1}, k_3=l+p_{12}, k_4=k+p_{123}, k_5=k-l. 
\end{equation}
The kinematics of the external momenta $p_i$ that we choose are: 
\begin{equation}
\sum_{i=1}^4 p_i =0,  \quad p_1^2=p_2^2=0, \quad p_{12}^2=p_{34}^2=s\, , \quad
p_{23}^2=p_{14}^2=t\, , 
\end{equation}
where $p_3^2$ and $p_4^2$ may or may not be on-shell.   

The bubble-box integral possesses an ultraviolet singularity due to a one-loop (bubble) subgraph. 
In addition to the ultraviolet divergence, we encounter one ``single-soft''  ($S_2$, in the notation of the previous section) and two ``single-collinear'' ($C_{k_1 || p_1}$, $C_{k_3 || p_2}$) singularities. 
When momentum $p_4$ is lightlike, the diagram also has a ``two-loop collinear" pinch surface, where both
loop momenta are parallel to $p_4$ and lines $k_1$, $k_4$ and $k_5$, share the momentum $p_4$, all moving toward the final state.    A similar pinch surface arises when $p_3$ is lightlike.   It is easy to check from Eq.\ (\ref{eq:power-ctg-scalar}), however, that, because they involve two collinear loops and only three collinear lines, these
pinch surfaces do not produce a singularity in the integral in four dimensions.   Therefore, it is not necessary to introduce counterterms for two-loop collinear pinch surfaces of this diagram in four dimensions, although in general it would be necessary in three dimensions.

We subtract approximations for the soft and 
collinear singularities sequentially, which automatically eliminates double-counting. As above, this leads to a remainder that is free of infrared singularities.
%
We implement collinear counterterms in the same manner as in the case of the diagonal box, explicitly,
\begin{eqnarray}
\label{eq:bubblebox_fin}
\left. \Bbox\right|_{\rm fin} &=&\
\left( 1 \, -\, t_{S_2} \, -\, t_{C_1}\left[ 1\, -\, t_{S_2}\right] \, -\, t_{C_3}\left[ 1\, -\, t_{S_2}\right] \right)\, \Bbox
\nonumber \\[2mm]
&=&\ \int \frac{d^dk_2}{i \pi^\dhalf}
\frac{d^dk_5}{i\pi^\dhalf}  
\left\{ 
\frac{1}{A_1 A_2 A_3 A_4  A_5} 
- \frac{1}{A_1 A_2 A_3} \left[  \frac{1}{A_4 A_5}\right]_{k_2=0} 
\right. 
\nonumber \\ 
&& 
\left.  
-\left[ 
\frac{1}{A_1 A_2} - \frac{1}{B_1 B_2}
\right]
\frac{1}{ s (1-x_1)}\left[  
\left. \frac{1}{A_4 A_5} \right|_{k_1 =-x_1 p_1} 
-
\left. \frac{1}{ A_4 A_5} \right|_{k_1 =- p_1} 
\right]
\right. 
\nonumber \\ 
&& 
\left. 
-\left[\frac{1}{A_2 A_3} - \frac{1}{B_2 B_3}\right] \frac{1}{ s (1-x_2)}\left[  
\left. \frac{1}{A_4 A_5} \right|_{k_3 =x_2 p_2} 
-
\left. \frac{1}{ A_4 A_5} \right|_{k_3 =p_2} 
\right]
\right\} \, ,
\nonumber\\
\end{eqnarray}
where $B_i \equiv A_i -\mu^2$. 

The above subtractions suffice to render the integral finite also in the ultraviolet limit. 
Indeed, the original integrand and the soft  counter-term in the first line have the same behaviour in the ultraviolet and cancel each other in that limit.  Given that the bubble-box integral is known analytically 
\cite{Anastasiou:1999bn}
and  that we can integrate simply the soft and collinear counterterms, it is
straightforward to check that indeed the remainder of
Eq.~(\ref{eq:bubblebox_fin}) has no $1/\epsilon$ poles and is
finite in $d=4$ dimensions.  
Specifically, for $t=-y \, s, $ with $0 \leq y \leq 1$, we find that 
\begin{eqnarray}
\label{eq:bubblebox_fin_res}
 \left. \Bbox\right|_{\rm fin} &=&\, 
-S_{12}(1-y)-3\,\zeta_3-\frac{\pi^2}{3}\,\log\left(\frac{\mu^2}{s}
                                    \right)
+\frac{1}{6}\,\log(y)^3 \nonumber \\ 
&& + i  \pi \left[ 
{\rm Li}_2(1-y)-\frac{\pi^2}{6}+\frac{1}{2}\,\log(y)^2
\right] 
\end{eqnarray}
The logarithmic term here that depends on the scale $\mu$ originates from
the integration of the collinear counterterms. The  $S_{12}$ Nielsen polylogarithm is defined in the appendix~\ref{sec:polylogarithms}. 

\subsection{Subtraction for the two-loop  planar double-box integral}

\begin{figure}[h]
\begin{center}
\includegraphics[width=0.5\textwidth]{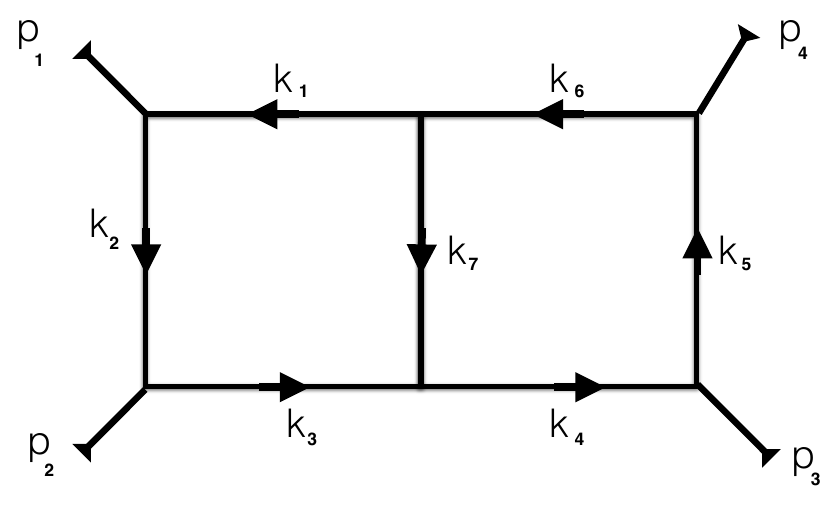}
\caption{\label{fig:planar-box} The two-loop double-box}
\end{center}
\end{figure}
We now consider the planar double-box two-loop integral, which was computed for the first time analytically in  Ref.~\cite{Smirnov:1999gc}. 
It is defined as
\begin{equation}
\label{eq:planarbox_definition}
\Pbox \equiv \int \frac{d^dk_2}{i \pi^\dhalf}
\frac{d^dk_5}{i\pi^\dhalf}  \frac{1}{A_1 A_2 A_3 A_4  A_5 A_6 A_7}, \quad 
\end{equation}
with 
\begin{equation}
A_i = k_i^2 + i0. 
\end{equation}
The momenta $k_i$ of the propagators are depicted in
Fig.~\ref{fig:planar-box}. One can concretely choose
\begin{equation}
k_1=l, k_2=l+p_1, k_3=l+p_{12}, k_4=k+p_{12}, k_5=k+p_{123}, k_6=k,
k_7=k-l. 
\end{equation}
The kinematics of the lightlike external momenta $p_i$, as usual defined to flow into the diagram, are taken to be
\begin{equation}
\sum_{i=1}^4 p_i =0,  \quad p_i^2=0, \quad p_{12}^2=p_{34}^2=s, \quad
p_{23}^2=p_{14}^2=t\, .
\end{equation}
For later use, we also define a generalisation of the scalar integral, with
an arbitrary numerator $N(k_2, k_5)$
\begin{equation}
\label{eq:planarbox_definition-sub}
\Pbox\left[ N\right] \equiv \int \frac{d^dl}{i \pi^\dhalf}
\frac{d^dk}{i\pi^\dhalf}  \frac{N(l, k)}{A_1 A_2 A_3 A_4  A_5 A_6 A_7}\, .
\end{equation}

The double-box with on-shell external lines has infrared
singularities, which are presented below, ordered according to
increasing volumes of their regions:  
\begin{itemize}
\item {\it Double-soft:} isolated points in both loop momentum spaces, where some line has a vanishing momentum for each loop.   We label these as $S_{i}S_{j}$, in which the internal
 lines $i,j$ become soft.
\item {\it Soft-Collinear:} an isolated point in one of the two loop spaces where one line has vanishing momentum, and a line segment of the other loop collinear to one of the external momenta.  These are labelled as $S_iC_{j||l}$, in which the internal particle  $i$ is soft and the internal particle $j$  is parallel to the
  external particle $l$.  Note that at these pinch surfaces there is always a two-fold ambiguity in the choice of collinear line $j$.
\item {\it Two-collinear pairs:} line segments for both loops, collinear to different external momenta, labelled $C_{i || l} C_{j ||m}$, in which the
  internal particles $i,l$ are parallel to the external particles
  $k,m$ respectively. \\
{\it Two-loop-collinear:} segments in which the two loops are both collinear to the same external momentum.   These are identified $C_{ijk||m}$, in which the internal
  particles $i,j,k$ are all  parallel to the external particle $m$.  Again, there is a trivial ambiguity in the choice of lines $i,j,k$, because four lines go on-shell.
\item  {\it Single-Soft:} an isolated point in one loop, and unconstrained momentum in the other, identified by $S_i$, in which the internal particle $i$ is
  soft.
\item {\it Single-Collinear:} a collinear segment in one loop, and unconstrained momentum in the other, identified by $C_{i||l}$ in which the internal
  particle $i$ is parallel to the external particle $l$.
\end{itemize}
Near each of these pinch surfaces, the integral is logarithmically divergent, as may be verified by applying power counting according to Eq.\ (\ref{eq:power-ctg-scalar}).   

As is characteristic of integrals with both collinear and infrared divergences, we expect both double and single poles from each loop integral, and indeed, these singularities yield up to  $1/\epsilon^4$ poles. 
To put our subtractions in context, we recall first the explicit form of the leading and next-to-leading poles of the planar box.
For $s>0,t<0$,
we have~\cite{Smirnov:1999gc} 
\begin{eqnarray}
\label{eq:Pbox-ldg}
\Pbox\left[ 1\right] = \frac{\Gamma^2(1+\epsilon)}{s^2 t} \left[ 
\frac{4}{\epsilon^4} + \frac{(3i\pi-5\ln(-t)-3\ln(s)}{\epsilon^3}
\right] +{\cal O}\left( \frac{1}{\epsilon^2}\right)\, .
\end{eqnarray} 

In contrast to the two-loop examples of the diagonal and bubble boxes above, a direct construction of all
possible nested subtractions, as in Eq.\ (\ref{eq:forest}) would involve nestings with as many as five approximation
operators and well over a hundred terms.   We will show below, however, that the IR singularities of the planar box, associated with the pinch surfaces listed above, can be subtracted with a much smaller set
of counterterms.    As in the previous examples, we construct these inductively, working from smaller to larger regions.  For each such region, we let the diagram, including subtractions for smaller regions, determine the subtraction for the region at hand.
We thus start the process with the double soft configurations (isolated points in both loops).   We design subtractions for these regions to cancel the leading poles.

\begin{figure}[h]
\begin{center}
\includegraphics[width=0.8\textwidth]{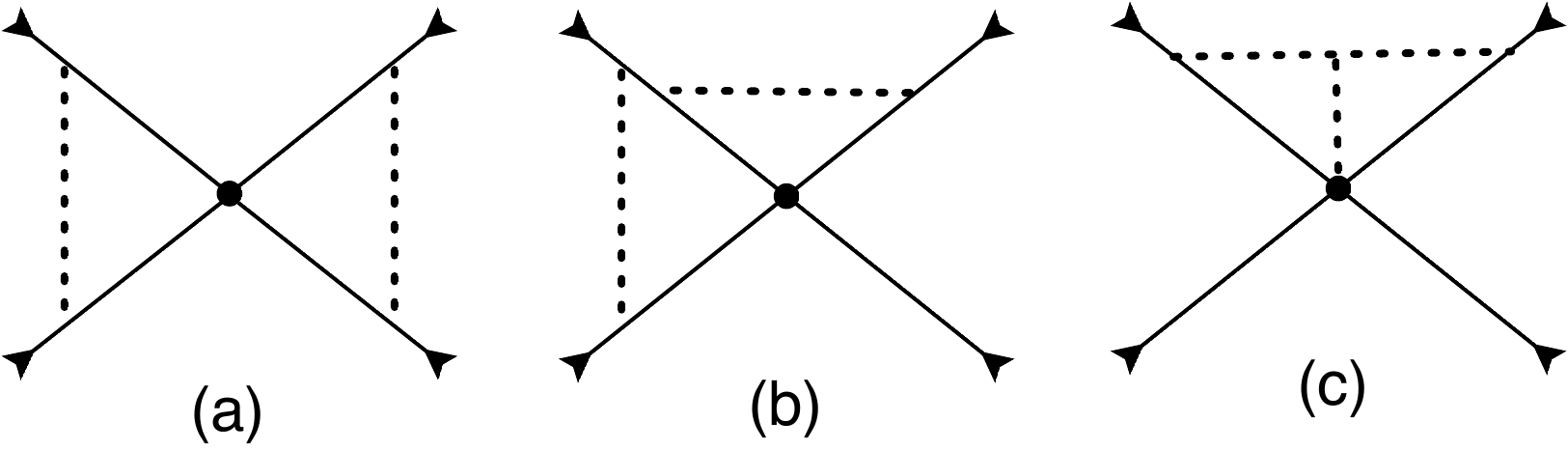}
\caption{\label{fig:double-soft} Representative reduced diagrams for double-soft pinches.   In (a) and (b), the denominator of single line is fixed at $t$ and $s$, respectively, while in (c) two lines are off-shell by $s$.}
\end{center}
\end{figure}

There are actually two types of double-soft points.   In the first class, illustrated by Fig.\ \ref{fig:double-soft}(a,b), exactly two
lines have vanishing momenta, while in the second class, illustrated by Fig.\ \ref{fig:double-soft}(c), three lines are joined together with zero momentum.   Notice that in the latter class, one of the soft lines attaches to the ``hard part" of the diagram, the two lines that are off-shell by invariant $s$ at the pinch surface.   A configuration like this is never leading in a gauge theory diagram, where the three-boson vertex is associated with a momentum factor, but it has power counting zero for $\phi^3$ in four dimensions, and so must be included in the analysis of the planar double box.

We choose to subtract first for the former class, consisting of $S_2S_5, S_2S_7, S_5S_7, S_2S_4$,
$S_2S_6$, $S_5S_1$ and $S_5S_3$, all
double-soft divergences. In these limits,  as reflected in the reduced diagrams of Fig.\ \ref{fig:double-soft}(a) and (b),
six of the seven propagators are on-shell and 
one propagator is hard:   
\begin{eqnarray}
S_2S_5: &&  A_7 \sim t, \quad
 S_2S_7: \; A_5 \sim t, \quad
S_5S_7: \;  A_2 \sim t, \nonumber \\
S_2S_4: &&  A_6 \sim s,  \quad
S_2S_6: \; A_4 \sim s,  \quad
S_5 S_1: \; A_3 \sim s, \quad
S_5 S_3: \; A_1 \sim s\, . 
\label{eq:soft-one-off}
\end{eqnarray}

The subtractions are defined to simply replace each off-shell momentum by the appropriate invariant.   For example, 
in the $S_2S_5$ subtraction, this is accomplished by a factor $A_7/t$.   
We label the planar box subtracted for each of these singularities as  $\Pbox[N_1]$, with the numerator $N$ in
equation (\ref{eq:planarbox_definition-sub}) given by 
\begin{equation}
N_1(l,k)  = 1 - \frac{A_{257}}{t} -\frac{A_{1346}}{s}\, ,
\end{equation}
where again $A_{257}=A_2+A_5+A_7$ and so on.   It is easy (and important)
to check that the subtraction term associated with a given $S_iS_j$ is
power-counting finite in the regions around the other points $S_kS_l$ in the list of Eq.\ (\ref{eq:soft-one-off}).
The planar box $\Pbox[N_1]$ is therefore free of all the above seven double-soft
singularities.  We have not, however, yet dealt with the two singular points where five propagators are on shell, and
two lines are off shell.  We label these points as 
\begin{eqnarray}
S_1 S_6: && A_3, A_4 \sim s\, ,
\nonumber  \\
S_3 S_4: && A_1, A_6 \sim s\, .
\label{eq:soft-two-off}
\end{eqnarray}
The remaining effect of these regions can be found by direct calculation.
Indeed, the subtraction terms in $\Pbox[N_1]$ can be evaluated by standard methods, and combined with the
results for the full diagram, given at the first two powers in $1/\epsilon$ by Eq.\ (\ref{eq:Pbox-ldg}).  The complete result is
\begin{eqnarray}
\Pbox\left[ N_1\right] = \frac{\Gamma^2(1+\epsilon)}{s^2 t} \left[ 
-\frac{1}{2\epsilon^4} + \frac{\ln(-t)}{\epsilon^3}
\right] +{\cal O}\left( \frac{1}{\epsilon^2}\right)\, .
\label{eq:Pbov-N1}
\end{eqnarray} 
In this expression, leading powers are still present,
so we must certainly make further subtractions.    
To do so, we follow the approach mentioned above, and determine the necessary
counterterms by studying the behavior of the full subtracted diagram $\Pbox[N_1]$,
rather than that of the original diagram, $\Pbox[1]$.   We find that two
counterterms with factors of $1/s$ in $\Pbox[N_1]$ are singular in each of the regions of Eq.\ (\ref{eq:soft-two-off}).
The net result is that the regions of (\ref{eq:soft-two-off}) are actually already subtracted {\it twice} in
$\Pbox[N_1]$.   Counterterms must thus add these regions back, to avoid double counting, rather than to
make an additional subtraction.

We thus arrive at the next stage in the subtracted
planar box, labelled $\Pbox[N_2]$, with a numerator 
\begin{equation}
\label{eq:N_2}
N_2 = 1 - \frac{A_{257}}{t} -\frac{A_{1346}}{s} +\frac{A_1A_6+A_3 A_4}{s^2} \, . 
\end{equation}
This expression is free of all double-soft singularities. Upon integration including the counterterms, it gives to order $\epsilon^{-2}$ the
explicit form
\begin{eqnarray}
\Pbox\left[ N_2\right] = \frac{1}{s^2 t} \left[ 
\frac{
-\frac{\pi^2}{3} -3 \ln^2\left( -\frac t s\right) - 6 i \pi \ln\left( -\frac t s\right)
 }{\epsilon^2} 
\right] +{\cal O}\left( \frac{1}{\epsilon}\right)\, .
\end{eqnarray} 
As anticipated,  after the subtraction of the approximations to the
integrand in all double-soft singular limits, the $1/\epsilon^4$ pole
is cancelled.  Somewhat unexpectedly, the $1/\epsilon^3$ is also
cancelled.  Evidently, our double-soft counterterms have removed 
further divergences, beyond those for which they were originally defined.   In
any case, $\Pbox[N_2]$ will be the starting point for the next round in
the construction. 

Next in our ordering of subtractions of ``increasing volumes'' 
comes the subtraction of {\it soft-collinear} singularities, which may be labelled as
\[S_iC_{k_2||p_1}, i=5,6, \quad S_iC_{k_2||p_2}, i=4,5\, , \] 
and 
\[S_iC_{k_5||p_3}, i=2,3, \quad S_iC_{k_5||p_4}, i=1,2. \]
We find that in these eight, potentially singular, soft-collinear limits the numerator $N_2$, Eq.\
(\ref{eq:N_2}) that we have generated  after the subtraction of the double-soft
divergences vanishes:  
\begin{equation}
S_iC_j : N_2 \to 0. 
\end{equation}
Therefore, we do not need to add any further counterterms in $N_2$ to remove soft-collinear
divergences.  This result is clearly a reflection of the lack of $1/\epsilon^3$ 
poles in $\Pbox[N_2]$.    We are ready to move on to the remaining regions,
which can give at most a double pole.

We now proceed with the subtraction of divergences of two-collinear
pairs (in which, five propagators are on-shell). 
 We find that the numerator $N_2$ vanishes in the limits,   
\begin{eqnarray}
C_{k_2 || p_1} C_{k_5||p_4}, C_{k_2 || p_2} C_{k_5||p_3}: N_2 \to 0\, , 
\end{eqnarray}
but it is finite in the remaining two limits of this type,
\begin{eqnarray}
C_{k_2 || p_1} C_{k_5||p_3}:&&  N_2 \to -\frac{s+t}{s^2 t} A_3 A_6 \, ,
\nonumber \\
C_{k_2 || p_2} C_{k_5||p_4}:&&  N_2 \to -\frac{s+t}{s^2 t} A_1 A_4 \, .
\end{eqnarray}
We therefore add two counterterms to $N_2$, which subtract the behavior in these limits,
\begin{equation}
N_3 = 1 - \frac{A_{257}}{t} -
             \frac{A_{1346}}{s} 
             +\frac{A_1A_6+A_3 A_4}{s^2} 
+\frac{s+t}{s^2 t} \left( A_1 A_4  + A_3 A_6\right)\, .
\end{equation}
The integral $\Pbox[N_3]$ is then free of divergences associated with {\it two-collinear pair}
singularities as well.   No new singularities from smaller regions are
produced, because every such singularity involves the vanishing of
at least one of the pair of denominators $A_1,A_4$ and $A_3,A_6$.

Of a similar order are singularities due to four {\it two-loop-collinear}
limits, in which $N_3$ scales as
\begin{eqnarray}
C_{k_2k_6k_7 || p_1} : &&  N_3 \to  -\frac{A_3 A_5}{st} , \quad 
C_{k_2k_4k_7 || p_2} : \;  N_3 \to  -\frac{A_1 A_5}{st} , \nonumber \\
C_{k_5k_3k_7 || p_3} : &&  N_3 \to  -\frac{A_6 A_2}{st} , \quad 
C_{k_5k_1k_7 || p_4} : \;  N_3 \to  -\frac{A_4 A_2}{st} \, .
\end{eqnarray}
To remove them, we add four additional (positive) counterterms to $N_3$, leading to 
\begin{eqnarray}
\label{eq:N-4}
N_4 &=&  1 - \frac{A_{257}}{t} -
             \frac{A_{1346}}{s} 
             +\frac{A_1A_6+A_3 A_4}{s^2} 
             +\frac{A_{13} A_5 + A_{46} A_2 }{st}  \nonumber \\ 
&&   \qquad 
+\frac{s+t}{s^2 t} \left( A_1 A_4  + A_3 A_6\right)\, .
\end{eqnarray}
The integral $P[N_4]$ is free of all singularities due to soft/collinear configurations that
involve both loop momenta, although on evaluation, it still retains double poles,
\begin{eqnarray}
\Pbox\left[ N_4\right] = \frac{1}{s^2 t} \left[ 
\frac{
-2 \ln^2\left( -\frac t s\right) - 4 i \pi \ln\left( -\frac t s\right)
 }{\epsilon^2} 
\right] +{\cal O}\left( \frac{1}{\epsilon}\right)\, .
\end{eqnarray} 
This is because the resulting integrand,
\begin{equation}
\label{eq:F_Pbox2}
F_{Pbox}^{(2)} \equiv 
\frac{N_4}{A_1 A_2 A_3 A_4 A_5 A_6 A_7},  
\end{equation}
is still divergent from pinch surfaces in which a single momentum is soft and/or collinear to an
external momentum.    
To remove these, we will use insight from the example of the one-loop box
treated in the previous section, turning our attention to the full integrand.

We deal first with the
singularities in the single-soft $S_2$ and $S_5$ limits. In those, the
integrand is approximated by  
\begin{equation}
S: F_{Pbox}^{(2)} \sim - F_{Pbox}^{(1s)}
\end{equation}
where the resulting counterterms are given by
\begin{eqnarray}
\label{eq:F_Pbox1s}
F_{Pbox}^{(1s)} &=& -\frac{1}{A_1 A_2 A_3} \left[  
\frac{
N_4
}
{
A_4 A_5 A_6 A_7
}
\right]_{k_2=0} 
-
\frac{1}{A_4 A_5 A_6} \left[  
\frac{
N_4
}
{A_1 A_2 A_3 A_7}
\right]_{k_5=0}.
\end{eqnarray}
The counterterms in $F_{Pbox}^{(1s)}$ due to single-soft singular
limits  are actually straightforward to integrate. 
For example, the single-soft counterterm
\begin{equation}
\frac{1}{A_1 A_2 A_3} \left[  
\frac{
N_4
}
{
A_4 A_5 A_6 A_7
}
\right]_{k_2=0}, 
\end{equation}
describes a one-loop subgraph with the propagators $A_4,\, A_5,\, A_6$ and  $A_7$, which does not contain the soft propagator, and is
evaluated at a fixed value $k_2=0$. This
operation factorizes the counterterm into the  product of two one-loop
integrals,  a singular one-loop triangle (containing the propagators
$A_1,A_2,A_3$) and an one-loop box (containing the propagators
$A_4,A_5,A_6,A_7$) with a numerator found by setting $A_1,A_2$ and $A_3$ to zero in Eq.\ (\ref{eq:N-4}),
\begin{equation}
\label{eq:pbox_softsubgraph}
\left. N_4 \right|_{k_2=0} = 1 - \frac{A_{46}}{s} -\frac{A_{57}}{t}\, .
\end{equation}
This expression is precisely the full numerator
for the subtracted one-loop box, Eq.\ (\ref{eq:N-box}), which renders the $k_5$ integral fully finite in this case.
This is not an accident, but a consequence of the fact that singularities stronger
than single-soft have been removed earlier in our
sequence of nested subtractions.  

With the single-soft singularities subtracted, 
the sum of $F_{Pbox} ^{(2)} + F_{Pbox}^{(1s)}$, Eqs.\ (\ref{eq:F_Pbox2}) and (\ref{eq:F_Pbox1s}), should only have {\it
  single-collinear} singularities, which yield single $1/\epsilon$
poles.  Indeed, upon integration we find
\begin{eqnarray}
&& \int \frac{d^dk_2}{i \pi^\dhalf}
\frac{d^dk_5}{i\pi^\dhalf} 
\left[ F_{Pbox}^{(2)} +  F_{Pbox}^{(1s)}
\right] =\frac{1}{s^2 t \epsilon} 
\left[ - 8 S_{12}\left(1+\frac t s \right)
+\frac{4}{3} \ln^3\left( - \frac t s\right)
\right. 
\nonumber \\  && 
\hspace{1cm} 
\left. 
-\frac{4}{3} \pi^2 \ln\left( - \frac t s\right) 
+4 i \pi \left( 
\ln^2\left( - \frac t s \right) + 2 {\rm Li}_2\left(1 + \frac t s  \right)
\right)
\right] 
+{\cal O}\left(\epsilon^0 \right)\, .
\end{eqnarray}
Explicitly, the integrand of $F_{Pbox}^{(2)} +  F_{Pbox}^{(1s)}$ is
singular only in the $C_{k_2||p_1},C_{k_2||p_2},C_{k_5||p_3},C_{k_5||p_4}$ single collinear
limits. In these limits, we can approximate the integrand by our final counterterm, $F_{Pbox}^{(1c)}$ defined by
\begin{equation}
F_{Pbox}^{(2)} +  F_{Pbox}^{(1s)} \sim - F_{Pbox}^{(1c)}
\end{equation}
in the remaining collinear limits.
This counterterm is given by an expression which, although a little long, is a straightforward generalization of the UV-finite diagonal-loop box subtractions in Eq.\ (\ref{eq:diagb_fin}),
\begin{eqnarray}
\label{eq:F-1c}
  F_{Pbox}^{(1c)} &=& - \left[ \frac 1 {A_1 A_2} - \frac 1
                      {B_1 B_2}\right]\frac{1}{ s (1-x_1)} \left\{
                      \left[ \frac{N_4}{ A_4 A_5 A_6 A_7} \right]_{k_1=-x_1 p_1} -
                      \left[\frac{N_4}{ A_4 A_5 A_6 A_7} \right]_{k_2=0} 
                      \right\}
                      \nonumber \\ 
                  && \hspace{-1.25cm}
- 
\left[ \frac 1 {A_2 A_3} - \frac 1
                      {B_2 B_3}\right]
\frac{1}{ s (1-x_2)} \left\{
                      \left[ \frac{N_4}{ A_4 A_5 A_6 A_7} \right]_{k_3=-x_2 p_2} -
                      \left[\frac{N_4}{ A_4 A_5 A_6 A_7} \right]_{k_2=0} 
                      \right\}
                      \nonumber \\ 
                  && \hspace{-1.25cm}
- \left[ \frac 1 {A_5 A_6} - \frac 1
                      {B_5 B_6}\right]
\frac{1}{ s (1-x_4)} \left\{
                      \left[ \frac{N_4}{ A_1 A_2 A_3 A_7} \right]_{k_6=x_4 p_4} -
                      \left[\frac{N_4}{ A_1 A_2 A_3 A_7} \right]_{k_5=0} 
                      \right\}
                      \nonumber \\ 
                  && \hspace{-1.25cm}
-\left[ \frac 1 {A_4 A_5} - \frac 1
                      {B_4 B_5}\right]
\frac{1}{ s (1-x_3)} \left\{
                      \left[ \frac{N_4}{ A_1 A_2 A_3 A_7} \right]_{k_4=-x_3 p_3} -
                      \left[\frac{N_4}{ A_1 A_2 A_3 A_7} \right]_{k_5=0} 
                      \right\},
\end{eqnarray}
with $B_i \equiv A_i -\mu^2$. 
In the above, collinear counterterms have been introduced in the same manner
as for the regulated diagonal-box, Eq.\ (\ref{eq:diagb_fin}), with fractional momenta defined by
\begin{equation}
x_1 = - \frac{k_1 \cdot \eta_1}{p_1 \cdot \eta_1}, \ \
x_2 = - \frac{k_3 \cdot \eta_2}{p_2 \cdot \eta_2}, \ \
x_3 = - \frac{k_4 \cdot \eta_3}{p_3 \cdot \eta_3}, \ \
x_4 =  \frac{k_4 \cdot \eta_4}{p_4 \cdot \eta_4}. 
\end{equation}
We have now specified our final counterterm and the integrand 
\begin{equation}
\label{eq:pbox_remainder}
F_{Pbox} = F_{Pbox} ^{(2)} + F_{Pbox}^{(1s)} + F_{Pbox}^{(1c)}, 
\end{equation} 
is free of all singularities.  

We have checked that as an analytic expression, the integral of
Eq.~(\ref{eq:pbox_remainder}) is finite in four dimensions, 
\begin{equation}
\label{eq:pbox_remainder_integ}
\Pbox{}^{\rm fin}\ =\ \int \frac{d^dk_2}{i \pi^\dhalf}
\frac{d^dk_5}{i\pi^\dhalf}  F_{Pbox} = {\cal O} \left( \epsilon^0 \right).
\end{equation} 
We have performed the analytic integration of the
counterterms in Eq.~(\ref{eq:pbox_remainder_integ}) in a straightforward
manner as we will explain shortly.  

To be specific, for $s>0$ and $0< y \equiv -\frac{t}{s} <1$  we find 
\begin{eqnarray}
\label{eq:Pboxfin_polylog}
s^2 t \Pbox{}^{\rm fin}\ =\ \left(
  C_R(y)  + i \pi C_I(y) \right) \log\left( \frac{\mu^2}{s} \right)
+A_R(y)  + i \pi A_I(y), 
\end{eqnarray}
with
\begin{eqnarray}
A_R(y) &=& 
-\frac 1 2 \,\log(y)^4-3\,\pi^2\,\log(y)^2+\frac{11
           \pi^4}{90}-\frac{14 \pi^2}{3}\,{\rm Li}_2(1-y) -24\,\zeta_3\,\log(y)
\nonumber \\ && 
+16\,\log(y)\,S_{12}(1-y)+32\,S_{13}(1-y)-12\,S_{22}(1-y)\, ,
\end{eqnarray}
\begin{eqnarray}
A_I(y) &=& 
-\frac{2}{3}\,\log(y)^3-\frac{2 \pi^2}{3}\,\log(y)-16\,\log(y)\,{\rm
           Li}_2(1-y)-24\,S_{12}(1-y)
\nonumber \\ && 
+12\,{\rm Li}_3(1-y)-4\,\zeta_3\, ,
\end{eqnarray},
\begin{eqnarray}
C_R(y) &=& -\frac{4} {3 }\,\log(y) \,\pi^2+\frac{4}{3}\,\log(y)^3-8\,S_{12}(1-y)\, ,
\end{eqnarray}
and
\begin{eqnarray}
C_I(y) &=& 
4\,\log(y)^2+8\,{\rm Li}_2(1-y)\, .
\end{eqnarray}
The finiteness of this result confirms that the counterterms reproduce all singular behavior of the analytic expression of the
planar double-box $\Pbox\left[1\right]$, as given in
Ref.~\cite{Smirnov:1999gc}. 

As anticipated, the counterterms are simpler than the full integral.
As we have noted above, the single-soft counterterms
reduce to the product of two one-loop integrals. The counterterms in $F_{Pbox}^{(2)}$,  due to double-soft,
soft-collinear, double-collinear pairs and two-loop-collinear singularities
contain two-loop integrals with at most six propagators.   We could reduce all
six-propagator integrals, following the algorithm of Ref.~\cite{Anastasiou:1999bn},
to the diagonal-box and bubble-box master integrals, which are simpler
than the original double-box integral and whose evaluation we have
described earlier.  
We anticipate that counterterms for such singularities (double-soft,
soft-collinear, two loop collinear and double-collinear-pairs) of an arbitrary
two-loop integral can be  expressed in terms of integrals with at most six
propagators, since this is the maximum number of propagators 
that can become on-shell in such configurations. 

The  integration of the counterterms in $F_{Pbox}^{(1c)}$, Eq.\ (\ref{eq:F-1c}), due to single-collinear
 limits  is slightly more involved, since it requires a one-dimensional convolution as described in 
 Eq.~(\ref{eq:collconv_result}) with a kernel that is an one-loop
 subdiagram.  We first reduce the one-loop subdiagram to the one-loop
 off-shell box and bubble master integrals of the
 appendix~\ref{sec:collmasters}. The remaining one-dimensional 
integrations yield Nielsen polylogarithms $S_{np}(y)$
(appendix~\ref{sec:polylogarithms})  of  uniform weight  $n+p \leq 4$. 
We have calculated the required integrals algebraically by comparing a few
first terms in their series expansion around $y=0, 1, \pm \infty$ 
and a general ansatz of such Nielsen polylogarithms.  

While here we studied a single two-loop diagram,  in a calculation of a physical two-loop amplitude it
is anticipated that sums of collinear limits from all diagrams
will factorize in terms of splitting functions times one-loop or tree
amplitudes, simplifying the convolutions into products.

\subsection{Subtraction for the two-loop crossed double-box integral}

\begin{figure}[h]
\begin{center}
\includegraphics[width=0.5\textwidth]{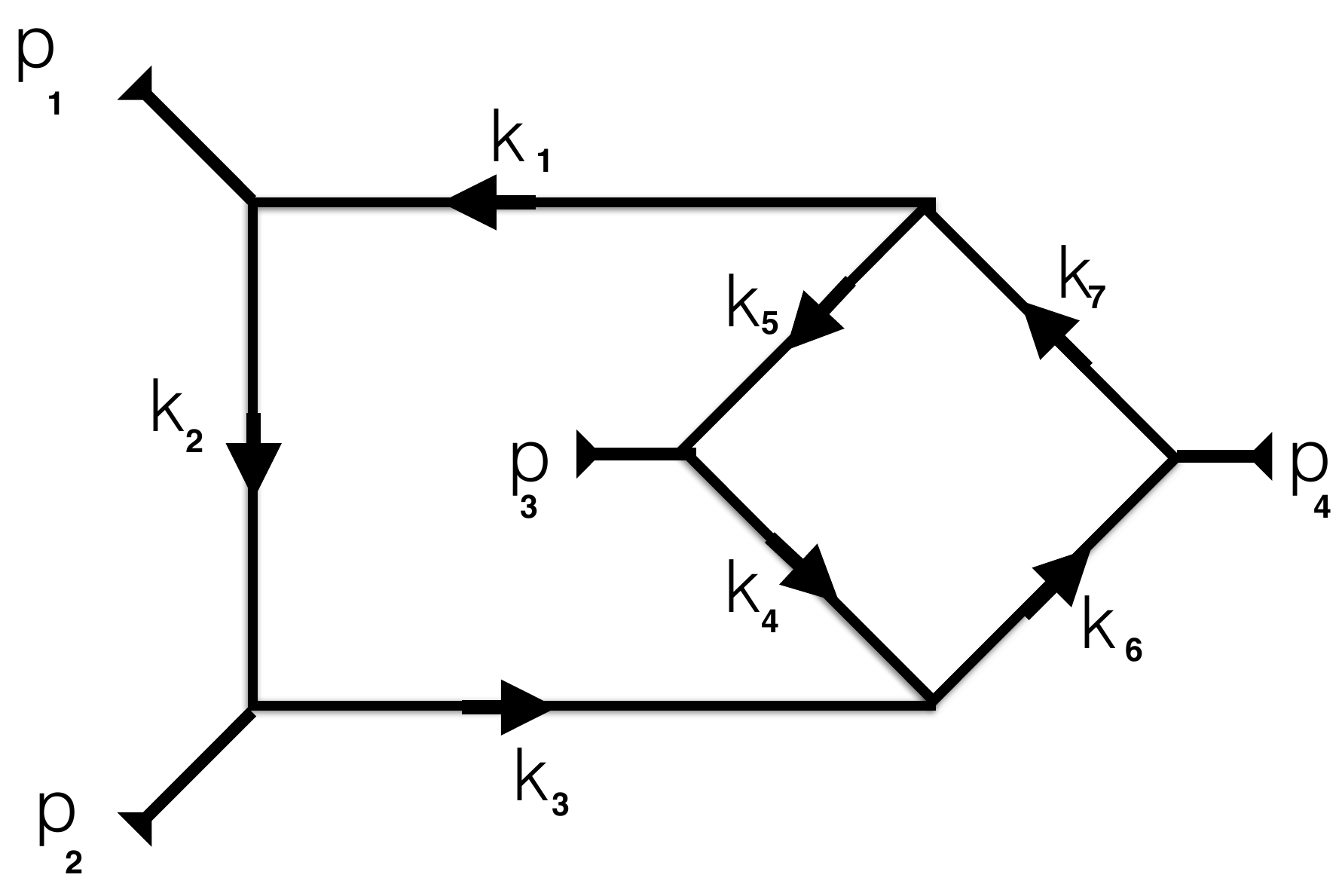}
\caption{\label{fig:cross-box} The two-loop cross-box}
\end{center}
\end{figure}

We now detail the construction of local counterterms for the two-loop
crossed double-box,
which is depicted in Fg.~\ref{fig:cross-box}. The
external momenta satisfy, 
\begin{equation}
p_1+p_2+p_3+p_4=0, \quad p_i^2=0, \quad p_{12}^2=s, \quad p_{23}^2=t,
\quad p_{13}^2=u=-s-t.  
\end{equation}
For convenience below, and as for the planar box, we introduce the integral with an arbitrary numerator $N$, and define 
\begin{eqnarray}
 \Xbox \left[ N\right] &\equiv& \int \frac{d^dk_2}{i \pi^\dhalf}
\frac{d^dk_5}{i\pi^\dhalf}  \frac{N(k_2, k_5)}{A_1 A_2 A_3 A_4  A_5 A_6 A_7},
\end{eqnarray}
with $A_i=k_i^2+i 0$. The internal momenta can be chosen as: 
\begin{eqnarray}
&& k_1=k,\ \  k_2=k+p_1, \ \ k_3=k+p_{12}, \ \ k_4 =-l-p_{12},  
   \nonumber \\  
&& k_5=-l+p_{4}, \ \ k_6=k-l, \ \ k_7=k-l+p_{4}. 
\end{eqnarray}
We are interested in removing the infrared singularities of $\Xbox[1]$,
which was computed analytically for the first time in
Ref.\ \cite{Tausk:1999vh}. 
We follow the same procedure as for the planar
double-box and previous examples. Namely, we 
remove the singularities iteratively, following
the order: double-soft, soft-collinear, two-collinear
pairs/two-loop-collinear, single-soft and single-collinear. 

\begin{figure}[h]
\begin{center}
\includegraphics[width=0.8\textwidth]{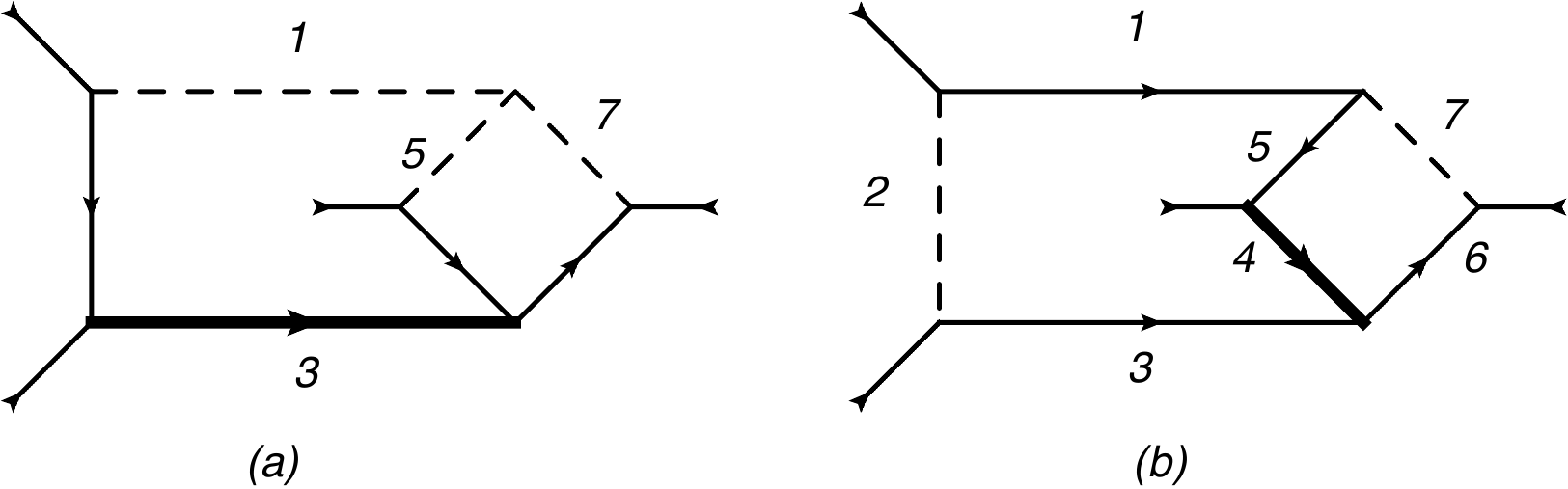}
\caption{\label{fig:cross-soft}  (a) Representation of the surface $S_1S_7$, at which the dashed lines
lines $k_1$, $k_7$ and $k_5$ have zero momentum, and at which the heavy line, $A_3=s$.  Other internal lines
carry a momentum equal to that of an external line.  (b) Representation of a typical
double-soft region at which two lines are soft. }
\end{center}
\end{figure}

Of the sixteen distinguishable double-soft regions of the crossed box, two
have the property that three lines are forced to zero momentum.   In the spirit of
our discussion for the planar box, we can
label these zero-dimensional pinch surfaces by any two of the three lines that 
are coupled at a three-point vertex and have vanishing momentum.
We will call them $S_1S_7$ and $S_3S_6$, where we understand that these
two configurations imply as well that $S_5$ and $S_4$ carry vanishing momentum,
respectively.   The region $S_1S_7$ is illustrated in Fig.\ \ref{fig:cross-soft}a.

At configurations of the cross-box like this, we encounter an additional 
complication, due to the presence of power-like (rather than logarithmic) double-soft singularities, 
\begin{eqnarray}
S_{1}S_{7} &:& \frac{d^dk_2 d^d k_5}{A_1 A_2 A_3 A_4 A_5 A_6 A_7} \to 
 \frac{d^dk_2 d^d k_5}{A_1 A_2 s A_4 A_5 A_6 A_7} 
\sim {\cal O}\left( \frac{\delta^4 \delta^4 }{ \delta^2 \delta \delta^0
               \delta  \delta^2 \delta \delta^2} \right)
\sim {\cal O}\left(
               \frac 1 \delta \right) \, ,
\nonumber \\ 
S_{3}S_{6} &:&  \frac{d^dk_2 d^d k_5}{A_1 A_2 A_3 A_4 A_5 A_6 A_7} \to 
 \frac{d^dk_2 d^d k_5}{s A_2 A_3 A_4 A_5 A_6 A_7} 
\sim {\cal O}\left( \frac{\delta^4 \delta^4 }{ \delta^0 \delta \delta^1
               \delta^2  \delta \delta^2 \delta} \right)
\sim {\cal O}\left(
               \frac 1 \delta \right) \, .
\nonumber \\ &&
\end{eqnarray} 
In order to remove these power-like singularities, we must introduce
a counterterm for which the numerator scales as $\delta^2$ in the
corresponding limits.  We achieve this with 
\begin{equation}
N_1  = \left( 1 - \frac{A_{13}}{s} \right)^2 \, ,
\end{equation}
which covers both cases.  Notice that the integral of $N_1$ includes the original diagram.

All other double-soft singularities are logarithmic and  we can
proceed to subtract them in an analogous way as in the 
planar double-box example.  
However, a subtlety remains, due to the original 
power-like nature of the $S_{1}S_{7}, S_{3}S_{6}$ singularities. 
As we have seen for the planar box, if one encounters only logarithmic singularities in the process of
performing nested subtractions, the integrands of counterterms that are introduced
later for 
higher-dimensional pinch surfaces
vanish in all limits of 
the lower-dimensional pinch surfaces
 that have been treated  earlier. In other words, once 
a logarithmic singularity is removed at a certain step it 
is not
 introduced
back spuriously with the construction of a subsequent counterterm for a different singularity. 
This mechanism is not automatic in the presence of a
power-like singularity, which  can influence  the construction of
counterterms for divergences which at first sight seem
unrelated to it.  

To illustrate this issue, let us consider the $S_2S_7$ log-like
singularity, in which:  
\begin{equation}
S_2S_7: \frac{N_1}{A_4} \to \frac{1}{u}. 
\end{equation}
To remove the $S_2S_7$ singularity,  we could add a counterterm to
$N_1$, such as 
\begin{equation}
N_2^\prime = N_1 -\frac{A_4}{u}. 
\end{equation}
However, this new counterterm has spoiled the cancelation of the
$S_1S_7$ limit which was achieved at the first step with $N_1$.
Indeed, 
\begin{equation}
  \frac{d^4 k_2 d^4 k_5 \; A_4}{A_1 A_2 A_3 A_4 A_5 A_6 A_7} \Big |_{S_1S_7}
\sim {\cal O}\left(\delta^0\right)\, . 
\end{equation}
Note that there is no such problem for the region of the 
$S_3S_6$ power-like singularity, where
 \begin{equation}
  \frac{d^4 k_2 d^4 k_5 \; A_4}{A_1 A_2 A_3 A_4 A_5 A_6 A_7} \Big |_{S_3S_6}\
\sim\ {\cal O}\left( \delta \right)\, .
\end{equation}
We can, however, easily construct a suitable counterterm that removes the
$S_2S_7$ singularity without introducing a spurious $S_1S_7$
singularity. Such a counterterm is found in the numerator: 
\begin{equation}
N_2 = N_1 -\frac{A_4}{u} \left( 1 - \frac{A_3}{s}\right) \, .
\end{equation} 
The double-soft $S_2 S_4, S_2 S_5$ and  $S_2S_6$ regions can be treated
similarly.  
Counterterms that cancel this set of singularities are given by $N_3-1$, where
\begin{eqnarray}
N_3  &=& \left(1 -\frac{A_{13}}{s} \right)^2-\left(1 -\frac{A_1}{s} \right) \left( \frac{A_5}{t} +
                \frac{A_7}{u}\right)
-\left(1 -\frac{A_3}{s} \right) \left( \frac{A_4}{u} +
                \frac{A_6}{t}\right).
\end{eqnarray}
There are two more double-soft singular limits that we have not
treated so far, $S_4S_7$ and $S_5S_6$. These limits leave the {\it same} propagator
($A_2$) hard, and set all other propagators on-shell,
\begin{eqnarray}
S_4 S_7&:& A_2 \to     u,  \quad N_3 \to 1, 
\nonumber \\  
S_5 S_6&:& A_2 \to t, \quad N_3 \to 1.  
\end{eqnarray}
Now that we only have one hard propagator at our disposal, 
no single fraction with $A_2$ in the numerator divided by a single invariant will serve
to subtract two different singularities.   Instead, we introduce a 
counterterm that interpolates between the values of $A_2$ at  the two singular configurations,
\begin{equation}
N_4^\prime = N_3 - \frac{A_2 (u+t-A_2)}{t u}=N_3 +\frac{A_2 (A_2+s)}{t u} \, .
\end{equation}
The new counterterm, $N_4^\prime$ vanishes in both $S_4S_7$ and $S_5S_6$ limits.
Unfortunately, it does not vanish fast enough in the $S_1S_7$ and $S_3
S_6$ limits where $\Xbox[1]$ develops power-like singularities.  
We have 
\begin{eqnarray}
S_1S_7&:& A_2 \sim \delta, A_1 \sim \delta^2,   A_3 \sim s   \nonumber \\ 
S_3S_6&:& A_2 \sim \delta, A_3 \sim \delta^2,   A_1 \sim s   \nonumber \\ 
\end{eqnarray}
In either of the above limits, $A_2 \sim \delta$ and only one of $A_1$ or $A_3$ tend
to the Mandelstam variable $s$.  We can therefore modify our counterterm
as follows:  
\begin{equation}
N_4 = N_3 +\frac{A_2 (A_2+s-A_{13})}{t u} 
\end{equation} 
The integral $\Xbox[N_4]$ is now free of all double-soft
singularities.  We also find that is free of all soft-collinear
singularities,
as confirmed by explicit integration.

We therefore proceed with the subtraction of
{\it two-collinear pairs/two-loop-collinear} types of
singularities. These singular limits do not pose any special
challenges and they are subtracted along the lines of our planar
double-box example. We find that the integral $\Xbox\left[
  N_5 \right]$ with numerator  
\begin{eqnarray}
N_5 &=& \left(1 -\frac{A_{13}}{s} \right)^2+\frac{A_2}{tu}
                   \left( A_2 + s  - A_{13}\right) 
\nonumber \\ &&
-\left(1 -\frac{A_1}{s} \right) \left( \frac{A_5}{t} +
                \frac{A_7}{u}\right)
-\left(1 -\frac{A_3}{s} \right) \left( \frac{A_4}{u} +
                \frac{A_6}{t}\right)
+\frac{A_2 A_{4567}}{t u} 
\nonumber \\ &&
-\frac{A_3}{s} \left(  \frac{A_7}{t} + \frac{A_5}{u}\right)
-\frac{A_1}{s} \left(  \frac{A_6}{u} + \frac{A_4}{t}\right)
+\frac{(t-u)^2}{s^2} \frac{A_1 A_3}{t u}
\end{eqnarray}
is free of all singularities 
associated with
two independent  loop momenta
pinched  
in a special
kinematic configuration (soft or collinear). 

Finally, we need to
remove the singularities due to single-soft and single-collinear
limits.  After these final subtractions, we find that the following
integrand is free of all singularities:
\begin{equation}
F_{Xbox} = F_{Xbox} ^{(2)} + F_{Xbox}^{(1s)} + F_{Xbox}^{(1c)}, 
\end{equation} 
where, following the notation of the planar double box,
\begin{equation}
F_{Xbox} ^{(2)} = \frac{N_5}{A_1 A_2 A_3 A_4 A_5 A_6 A_7},
\end{equation}
\begin{eqnarray}
F_{Xbox}^{(1s)} &=& -\frac{1}{A_1 A_2 A_3} \left[  
\frac{N_5}
{A_4 A_5 A_6 A_7}
\right]_{k_2=0} 
\end{eqnarray}
and
\begin{eqnarray}
F_{Xbox}^{(1c)} &=& -
\left[  
\frac{1}{A_1 A_2} -\frac{1}{B_1 B_2} 
\right]
\frac{1}{ s
                    (1-x_1)} 
\left\{ 
  \left[ \frac{N_5}{A_4 A_5 A_6 A_7} \right]_{k_1 =- x_1 p_1} 
-\left[ \frac{N_5}{A_4 A_5 A_6 A_7} \right]_{k_2 =0}
\right\}
\nonumber \\ 
&& 
\hspace{-1cm}
- \left[  
\frac{1}{A_2 A_3} -\frac{1}{B_2 B_3} 
\right]
\frac{1 }{s (1-x_3)} 
\left\{ 
\left[ 
\frac{N_5}{
                    A_4 A_5 A_6 A_7} \right]_{k_3 =- x_2 p_2} 
-\left[ 
\frac{
N_5
}{
                    A_4 A_5 A_6 A_7} \right]_{k_2 =0}
\right\}
\nonumber \\ 
&& 
\hspace{1cm}
-\left[  
\frac{1}{A_4 A_5} -\frac{1}{B_4 B_5} 
\right]
\left[\frac{
   N_5}{A_1 A_2 A_3 A_6 A_7}
   \right]_{k_5= - x_3 p_3} 
\nonumber \\ 
&& 
\hspace{1cm}
-\left[  
\frac{1}{A_6 A_7} -\frac{1}{B_6 B_7} 
\right]
\left[
   \frac{N_5}{A_1 A_2 A_3 A_4 A_5} \right]_{k_5= - x_4 p_4} \, .
\end{eqnarray}
In the above, $B_i  =  A_i - \mu^2$. 
Upon direct analytic integration, using the integration techniques described in
the previous section for the counterterms, and the analytic result of
\cite{Tausk:1999vh} for the crossed double-box integral, we verify that 
\begin{equation}
{\Xbox}^{\rm fin} \equiv \int \frac{d^dk_2}{i \pi^\dhalf}
\frac{d^dk_5}{i\pi^\dhalf}  F_{Xbox} = {\cal O}(\epsilon^0).
\end{equation}
Specifically, for $s>0$ and $ y \equiv -t/s \in [0, 1]$, we find 
\begin{equation}
s^3 {\Xbox}^{\rm fin} = \frac{f_{\Xbox}(y)}{y} + \frac{f_{\Xbox}(1-y)}{1-y}, 
\end{equation}
where 
\begin{equation}
\label{eq:Xboxfin_t_polylog}
f_{\Xbox}(y) = \left[ G_R(y) +i \pi G_I(y) \right] \, \log \left(
  \frac{\mu^2}{s}\right) + E_R(y) +i \pi E_I(y)
\end{equation}
and 
\begin{eqnarray}
E_R(y) &=& 
-8\,\pi^2\,{\rm Li}_2(y)+8\,{\rm Li}_2(y)\,\log(1-y)^2-28\,\log(y)\,{\rm Li}_2(y)\,\log(1-y)
-18\,{\rm Li}_2(y)\,\log(y)^2
\nonumber \\ && \hspace{-1cm}
+44\,{\rm Li}_3(y)\,\log(1-y)+96\,{\rm Li}_3(y)\,\log(y)-188\,{\rm Li}_4(y)+\frac{17}{36}\,\pi^4
+\frac{1}{ 12}\,\log(1-y)^4
\nonumber \\ && \hspace{-1cm}
+7\,\log(y)\,\log(1-y)\,\pi^2-\frac{25}{6}\,\pi^2\,\log(1-y)^2-\frac{3}{2}\,\log(y)^2\,\pi^2+\log(y)\,\log(1-y)^3
\nonumber \\ && \hspace{-1cm}
+44\,S_{12}(y)\,\log(1-y)-52\,S_{12}(y)\,\log(y)+84\,S_{13}(y)+88\,S_{22}(y)-44\,\zeta_3\,\log(1-y)
\nonumber \\ && \hspace{-1cm}
-4\,\log(y)\,\zeta_3-\frac{1}{4}\,\log(y)^4+\log(y)^3\,\log(1-y)-\frac{9}{2}\,\log(y)^2\,\log(1-y)^2,
\nonumber \\ && \hspace{-1cm}
\end{eqnarray}
\begin{eqnarray}
E_I(y) &=& 
-40\,{\rm Li}_2(y)\,\log(1-y)-24\,{\rm Li}_2(y)\,\log(y)+64\,{\rm
           Li}_3(y)+\frac{8}{3}\,\pi^2\,\log(1-y)
\nonumber \\ && \hspace{-1cm}
-6\,\log(y)\,\pi^2-60\,S_{12}(y)+56\,\zeta_3-\frac{2}{3}\,\log(y)^3-10\,\log(1-y)^2\,\log(y)
\nonumber \\ && \hspace{-1cm}
+\frac{2}{3}\,\log(1-y)^3,
\end{eqnarray}
\begin{eqnarray}
G_R(y) &=& -12\,{\rm Li}_2(y)\,\log(y)+12\,{\rm
           Li}_3(y)+\frac{2}{3}\,\pi^2\,\log(1-y) -\frac{8}{3}\,\log(y)\,\pi^2
\nonumber \\ && \hspace{-1cm}
-8\,S_{12}(y)-4\,\zeta_3+\frac{2}{3}\,\log(y)^3-4\,\log(y)^2\,\log(1-y)+\frac{2}{3}\,\log(1-y)^3,
\end{eqnarray}
and
\begin{eqnarray}
G_I(y) &=& 
-4\,{\rm
           Li}_2(y)+\frac{10}{3}\,\pi^2+4\,\log(y)^2-8\,\log(y)\,\log(1-y)+2\,\log(1-y)^2.
\nonumber \\ 
&& 
\end{eqnarray}

The integration of the counterterms was performed using the same
techniques as in the case of the planar double-box.  A notable
difference occurred in the integration of the collinear
counterterms. In the case of the crossed double-box, integrals
which do not have a representation in terms of Nielsen polylogarithms
with a simple argument $S_{np}(y)$ emerge~\footnote{We thank
  F. Dulat, F. Moriello and A. Schweitzer for providing useful confirmation
of this point.}. However, we have observed that the linear combination
which is required in the collinear counterterm can be expressed in
terms of Nielsen polylogarithms in our simple basis $S_{np}(y)$.  
Specifically, we find that
\begin{eqnarray}
&& \int_{0}^1 \frac{dx}{x} 
\left[ 
S_{12}\left( 
\frac{(x-y)(x y-1) }{y (x-1)^2}
\right)
- 2 {\rm Li}_2\left( 
\frac{(x-y)(x y-1) }{y (x-1)^2}
\right) \log(1-x) -\zeta_3
\right]  
\nonumber \\ 
=
&& -\frac{1}{24}\,\log(y)^4-2\,{\rm
   Li}_2(y)^2+\frac{13}{45}\,\pi^4-{\rm Li}_2(y)\,\log(y)^2+4\,{\rm
   Li}_3(y)\,\log(y)
\nonumber \\ &&
-4\,\zeta_3\,\log(y)-\frac{4}{3}\,\pi^2\,{\rm Li}_2(y)-8\,{\rm
                Li}_4(y)+8 \, S_{22}(y).
\end{eqnarray}


\section{Small mass expansions}
\label{sec:smallmass}
In the previous section, we rendered finite integrals that
were computed in dimensional regularisation.
Dimensional regularisation, however, is not an essential
element; our method is in principle applicable to any
infrared regulator.  Infrared divergences can also be regulated by a small
mass parameter.  With mass regularisation, 
the integration over the  mass-divergent regions yields logarithms that  
become infinite in the massless limit.
The mass regulator can be artificial or physical. 
For example, the physical mass of the bottom-quark in processes for
the production of Higgs bosons acts as a regulator for some
of the infrared divergences. The logarithmic dependence of the
corresponding amplitudes is of a high phenomenological interest. 
In this section, we will use the method of nested subtractions in
order to derive simply the asymptotic behavior  of 
certain Feynman  integrals 
in a small-mass limit.   

Consider a loop integral, represented schematically as
\begin{equation}
I[f_m] = \int dk_i\,   f(k_i, m)\, ,
\end{equation}
which depends on a small mass-parameter $m$, appearing in 
denominators of the standard form, $k_i^2-m^2+i0$.  
If we take  the zero-mass limit, the integral develops new infrared
divergences, which are not present for finite values of the mass.  
In general, these appear as logarithmic corrections in the mass,
which result from regions where the values of denominators
are actually larger than the mass:  $m^2\le k_i^2 \le Q$, with $Q$
some scale fixed by the invariants.   For our discussion, all
invariants are of the same order.   For fixed-angle scattering this is
the case, and logarithmic integrals can be identified by the simple
power-counting rules described in Sec.\ \ref{sec:method}.    
At the same time, the integral can receive finite contributions
from regions where for one or more denominators $k_i^2={\cal O}(m^2).$
We would like to find a systematic method to isolate both
the logarithmic mass dependence, and contributions that
are finite for small, but nonzero mass.    

To this end, we follow the method of nested subtractions, and construct an 
approximation $f_{\rm approx}(k_i, m)$ of the integrand in all the
limits that become singular as $m \to 0$.  As indicated above,
these limits can be identified using power counting techniques.
 Then, we can  use these
approximations to construct counterterms, keeping those mass-dependent terms
that dominate each denominator in the region for which the counterterm is designed
to approximate the full integral.   Thus, in general, our counterterms retain
mass dependence,
\begin{equation}
\label{eq:massive-subtraction}
I[f_m] = \int  f_{\rm approx}(k_i, m)
+ \int \left[ f(k_i, m) -f_{\rm approx}(k_i, m) \right]\, .
\end{equation}
The  leading mass singularities are now found from the integral
over the approximated integrand in the first term of the right-hand side of this relation. 
Because we keep all leading mass dependence, finite terms
associated with momenta of the order of the mass may remain
in one or more of the counterterms included in $f_{\rm approx}$. 

The second term in Eq.\ (\ref{eq:massive-subtraction}) is well-defined in the $m=0$ limit,
and additional finite terms, including important kinematic dependence, appear in general in
\begin{equation}
\label{eq:general_mass_expansion}
I[f_m] = \int  f_{\rm approx}(k_i, m)
+ \int \left[ f(k_i, m) -f_{\rm approx}(k_i, m) \right]_{m \to 0} + {\cal O}(m)\, .
\end{equation}
It is important to note that a naive Taylor expansion of the second term in
Eq.~(\ref{eq:massive-subtraction}) beyond the leading order 
will not account for terms of ${\cal O}(m^1)$, which vanish as a power, but may be multiplied by logarithms.
Because we drop terms that are non-leading of order in ${\cal O}(m)$ in denominators,
we can only ensure the cancellation of the leading power of $m$ in the second term,  
and will in general miss contributions of the form $m \log(m)$. 
In the following, using these ideas, we will show how this approach can be used to derive the small-mass dependence 
of 
one- and two-loop integrals.

  \subsection{One-loop massive triangle}  
\begin{figure}[h]
\begin{center}
\includegraphics[width=0.5\textwidth]{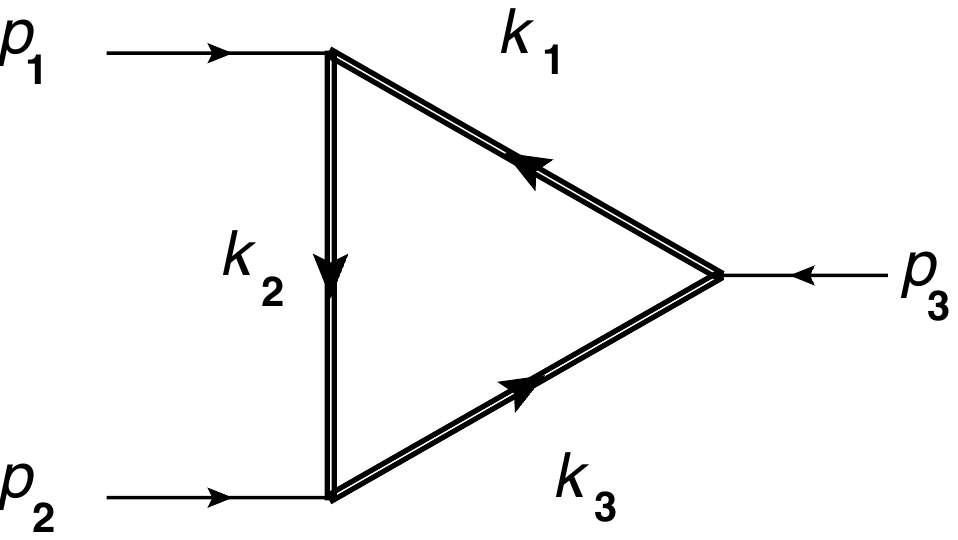}
\caption{\label{fig:massive1Ltriangle} The one-loop scalar triangle. Thick
  double lines denote massive propagators.}
\end{center}
\end{figure}


Consider as our first example the scalar one-loop massive triangle of
Fig.~\ref{fig:massive1Ltriangle}, taken with two light-like external lines, $p_1$ and $p_2$ here.   This is a rather  simple integral,
which hardly needs special treatment.   It is divergent in the zero-mass limit, however, and is useful to illustrate 
nested subtractions at finite mass.     We note that this analysis depends on having, as in
this case, only a single nonzero infrared-scale mass.  If, for example, we were to evaluate this diagram with one or more timelike
external line near the mass shell, the reasoning below would require further analysis.

The integral in question is straightforward to evaluate by standard methods and is given by   
\begin{equation}
\label{eq:I-tri-def}
I = \int \frac{d^4k_2}{i \pi^2}\frac{1}{B_1 B_2
   B_3}, 
\end{equation}
where the $B$'s are denominators with regulating masses,
\begin{equation}
B_1=k_1^2-m^2, \; B_2=k_2^2-m^2, \; B_3=k_3^2-m^2 \, .
\end{equation}
Referring to the figure, we choose $k_2$ as the loop momentum, so that the remaining momenta satisfy, with $p_1^2=p_2^2=0$,
\begin{equation}
k_1=k_2-p_1, \quad k_3=k_2+p_2, \quad p_1 \cdot p_2 \equiv \frac s 2\, .
\end{equation}
This integral is finite in $d=4$ dimensions but is logarithmically divergent as $m \to  0$: 
\begin{eqnarray}
\label{eq:massive_triangle}
 I &=& \frac{1}{2 s} \log^2\left(
\frac{\sqrt{1-\frac{4 m^2}{s}}-1}{\sqrt{1-\frac{4 m^2}{s}}+1}
+i0\right)^2 
\nonumber \\
&=&  \frac{1}{2 s} \left[ \log^2\left( \frac{m^2}{-s-i0} \right)  -4 \frac{m^2}{s} \log\left( \frac{m^2}{-s-i0} \right)
+{\cal O}\left( \frac{m^4}{s^2} \right)
\right] \, .
\end{eqnarray} 
To illustrate our method applied in four dimensions with regulator masses, we construct counterterms that isolate this
leading $m\to 0$ behavior, to the level of finite terms.    Of course, with an answer as simple as Eq.\ (\ref{eq:massive_triangle}), 
the counterterms will be at least as complex as the original diagram.   Our interest, however, is in the counterterm integrals themselves, because they may appear in more complex diagrams, where the asymptotic behavior of the full integral may be much more complex.

We now proceed to construct an integrand that is free of mass singularities.  As is well known, and easy to confirm, the vertex with two lightlike external lines has one soft and two collinear leading pinch surfaces.  The  singularity of lowest dimension corresponds to the the soft limit, and this is where we begin.  

In the massless limit, the soft singularity is associated with vanishing $k_2$.  In the presence of a mass regulator, the integrand is enhanced in the region
\begin{eqnarray}
\label{eq:soft-region-def}
S_2: k_2^\mu  &\sim& m, m \to 0\, ,
\nonumber\\
A_2 &\sim& m^2\, .
\end{eqnarray}
In this limit,  
\begin{equation}
A_1 =  k_1^2 -m^2 \sim - 2 k_2 \cdot p_1 \sim k_1^2, \;   A_3 =  k_3^2 -m^2 \sim 2 k_2 \cdot p_2 \sim k_3^2.
\end{equation}
We thus study the effect of a counterterm for this soft limit, subtracted from the original diagram, (\ref{eq:I-tri-def}), at the level of the integrand, 
\begin{eqnarray}
\label{eq:IRS-def}
I_{R_S} &\equiv& \int \frac{d^4k_2}{i \pi^2}\frac{1}{k_2^2-m^2} \left[ 
\frac{1}{k_1^2-m^2} \frac{1}{k_3^2-m^2}
-
\frac{1}{k_1^2} \frac{1}{k_3^2}
\right] \, .
\end{eqnarray}
In defining the subtraction, we have used that, for the soft region defined as in Eq.\ (\ref{eq:soft-region-def}), the invariant masses of the ``collinear" lines, $k_1$ and $k_3$ scale as $\sqrt{mQ}$, with $Q\sim\sqrt{s}$, so that $k_i^2\gg m^2$, $i=1,3$.   The counterterm is thus a good approximation to the full integral in the entire soft region.   In the soft region, it reproduces both logarithmic $m$ dependence and finite remainders.   Because we are calculating in a superrenormalizable theory, the behavior from infrared regions gives in fact the full leading-power behavior of the diagram.

This would be the entire story if the soft configuration were the only pinch surface of the massless triangle diagram.   As we know, however, it is not, and both the full integral and the counterterm as defined in (\ref{eq:IRS-def}) have pinches when the loop momentum $k_2$ is collinear to the external momentum $p_1$ or $p_2$.  The diagram and its soft counterterm are not guaranteed to behave exactly the same in this region.   We thus expect in general a leading-power ($m^0$) contribution to $I_{R_S}$, and indeed, by evaluating the integral that defines $I_{R_S}$ in (\ref{eq:IRS-def}), we find
\begin{equation}
\label{eq:IRS-zeta}
I_{R_S}\ = \ \frac{2}{s}\, \zeta(2)\, .
\end{equation}
Evidently, the soft subtraction generates the logarithmic $m$-dependence of the full integral, but cannot reproduce its finite part part without aid of collinear subtractions.   

The collinear pinch surface for $k_2$ aligned with $p_1$ is conveniently parameterized in the light-cone form,
\begin{equation}
k_2  = x_1 p_1 + \beta_1\eta_1 + k_{2 \perp}\, ,
\end{equation}
with
\begin{equation}
\label{eq:CO-1-tri}
x_1 =\frac{2 k_2 \cdot \eta_1}{ 2 p_1 \cdot \eta_1}  \sim {\cal O}(1), \; 
\beta_1 =\frac{2 k_2 \cdot p_1}{ 2 p_1 \cdot \eta_1}  \sim m^2, \;
k_{2 \perp} \sim m, \; m \to 0\, .  
\end{equation}
Following our iterative procedure, the full, soft-subtracted triangle, Eq.\ (\ref{eq:IRS-def}) is used to define  the counterterm for the collinear  region $C_{k_2 || p_1}$, using the behavior of its integrand  in that region, which we may represent as
\begin{eqnarray}
 \left[ 
\frac{1}{k_1^2-m^2} \frac{1}{k_3^2-m^2}
-
\frac{1}{k_1^2} \frac{1}{k_3^2}
\right] \, \Bigg |_{C_{k_2 || p_1}}
\ &=&\ \frac{1}{k_2^2-m^2} \left[ 
\frac{1}{k_1^2-m^2} 
-
\frac{1}{k_1^2} 
\right] \frac{1}{s x_1} \, ,
\nonumber\\[2mm]
&=& 
\frac{1}{k_2^2-m^2} \left[ 
\frac{m^2}{(k_1^2-m^2)(k_1^2)} 
\right] \frac{1}{s x_1} \, ,
\end{eqnarray}
with $x_1$ defined as in Eq.\ (\ref{eq:CO-1-tri}).
Analogously, a collinear singularity appears in $I_{R_S}$ when $k_2 = x_2 p_2$ with $x_2 =\frac{2 k_2 \cdot \eta_2}{2 p_2 \cdot \eta_2}$  and $m \to 0$.   Note that although approximating $k_3^2=sx_1$ produces ultraviolet singularities in each of the two terms on the right-hand side of the first equality, these divergences cancel when the two terms are combined.   Thus, we may stay in four dimensions for the entire calculation, without modifying our collinear subtractions.

We now remove the collinear mass singularities by introducing to the
integrand of $I_{R_S}$  two collinear counterterms, and define
\begin{eqnarray}
\label{eq:IRSC-def}
I_{R_{SC}} &\equiv& \int \frac{d^4k_2}{i \pi^2}\frac{1}{k_2^2-m^2} 
\left\{ 
\left[ 
\frac{1}{k_1^2-m^2} \frac{1}{k_3^2-m^2}
-
\frac{1}{k_1^2} \frac{1}{k_3^2}
\right] 
 \right. 
\nonumber \\ && 
\left. 
-\frac{1}{s x_1} \frac{m^2}{k_1^2 (k_1^2-m^2)}
-\frac{1}{s x_3} \frac{m^2}{k_3^2 (k_3^2-m^2)}
\right\}\, .
\end{eqnarray}
It is not difficult to verify that each of the collinear counterterms integrates precisely to $(1/s)\, \zeta(2)$, so that their sum cancels the extra finite term in $I_{R_S}$, Eq.\ (\ref{eq:IRS-zeta}), and the fully-subtracted diagram,  Eq.\ (\ref{eq:IRSC-def})  is free of all $m^0$ contributions.

We can summarize these results by writing the diagram as the sum of counterterms plus the fully-subtracted integral.    
\begin{eqnarray}
I &=& \int \frac{d^4k_2}{i \pi^2}
\frac{1}{k_2^2-m^2}
\left\{ 
\frac{1}{k_1^2} \frac{1}{k_3^2}
+\frac{1}{s x_1} \frac{m^2}{k_1^2 (k_1^2-m^2)}
+\frac{1}{s x_3} \frac{m^2}{k_3^2 (k_3^2-m^2)}
\right\} \nonumber \\ 
&& 
+
\int \frac{d^4k_2}{i \pi^2}\frac{1}{k_2^2-m^2} 
\left\{ 
\left[ 
\frac{1}{k_1^2-m^2} \frac{1}{k_3^2-m^2}
-
\frac{1}{k_1^2} \frac{1}{k_3^2}
\right] 
\right. \nonumber \\ 
&& \left.    
-\frac{1}{s x_1} \frac{m^2}{k_1^2 (k_1^2-m^2)}
-\frac{1}{s x_3} \frac{m^2}{k_3^2 (k_3^2-m^2)}
\right\}_{m=0} 
+ {\cal O}(m^2) \, .
\end{eqnarray}
We have seen above that the sum of counterterms reproduces mass dependence of the original diagram up to terms that vanish as a power of the mass $m$.   By construction, the counterterms also approximate the full integrand in all regions that give leading-power dependence.    We may therefore take the limit $m\to 0$ inside the integrand in the second integral, where it manifestly vanishes, as anticipated in Eq.\ (\ref{eq:general_mass_expansion}).  Note that although the terms proportional to $m^2$ appear to be multiplied by power-divergent integrals, there is no ambiguity, because each of these two terms is the sum of collinear subtractions for the full diagram with its soft subtraction, which cancel independently in the $m\to 0$ limit.  Indeed, as we have seen, the zero-mass limit may be taken before or after the integrals.

   In summary, we may write for the triangle diagram,
\begin{eqnarray}
I &=& \int \frac{d^4k_2}{i \pi^2}
\frac{1}{(k_2^2-m^2)}
\left\{ 
\frac{1}{k_1^2} \frac{1}{k_3^2}
+\frac{1}{s x_1} \frac{m^2}{k_1^2 (k_1^2-m^2)}
+\frac{1}{s x_3} \frac{m^2}{k_3^2 (k_3^2-m^2)}
\right\} 
+ {\cal O}(m^2) 
\nonumber \\ &=& 
\frac{1}{2 s} \left[ \log^2\left( \frac{m^2}{s} \right)  
+{\cal O}\left( \frac{m^2}{s} \right)
\right]\, .
\end{eqnarray}
which agrees with the result of Eq.~(\ref{eq:massive_triangle}).   Our subtraction procedure has thus succeeded in reproducing the infrared behavior of this diagram, including constants, using a mass regulator.   We emphasize again that although in this example the subtractions provide only a roundabout derivation of a simple result, the same  relatively simple subtractions can appear in diagrams with many more lines.

\subsection{One-loop massive box with one off-shell leg}

\begin{figure}[h]
\begin{center}
\includegraphics[width=0.5\textwidth]{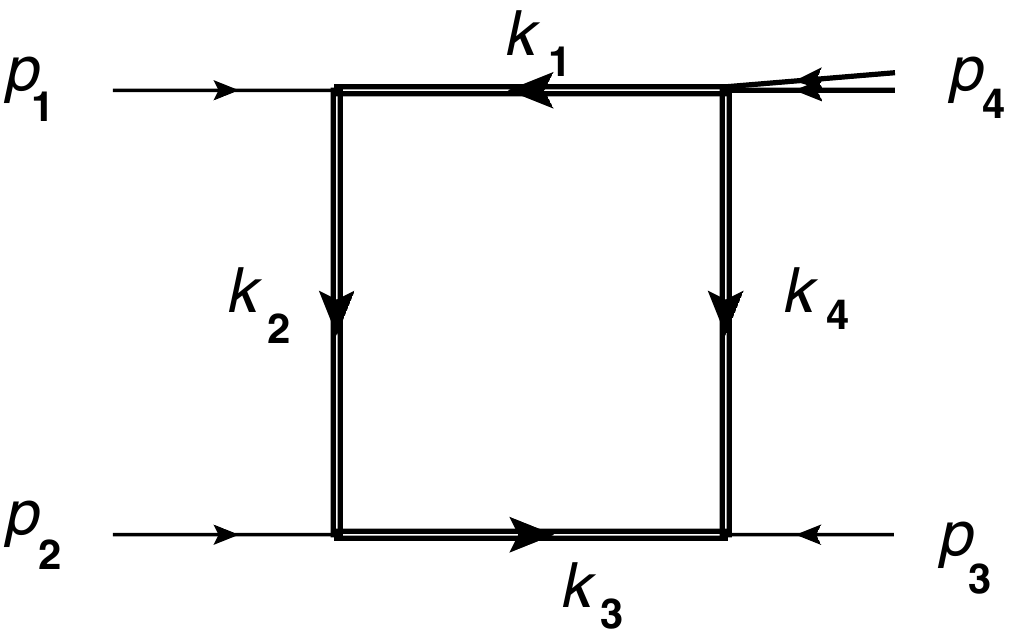}
\caption{\label{fig:massive1Lbox} The one-loop scalar box with
  one off-shell leg. Thick double lines denote massive propagators.}
\end{center}
\end{figure}
For our next example, we consider a slightly more complex integral.   This is the one-loop scalar-box integral with a single off-shell external line.    To anticipate, we will find it useful in this case to go back to dimensional regularization, but keeping the masses,
\begin{equation}
\label{eq:massive-box}
J = \int \frac{d^dk_2 }{i  \pi^\dhalf} \frac{1}{B_1 B_2 B_3 B_4}, 
\end{equation}
with 
\[ 
B_i =A_i -m^2, 
\]
where as usual
\begin{equation}
A_i^2\ =\ k_i^2 + i0\, .
\end{equation}
The momenta in the loop satisfy the conservation relations
\begin{equation}
k_2=k_1+p_1, \quad k_3=k_1+p_{12}, \quad  k_4 =k_1+p_{123}. 
\end{equation}
The external momenta satisfy $p_{1234}=0$ and 
\begin{equation}
p_1^2=p_2^2=p_3^2=0, \quad
p_{12}^2\equiv s, \quad p_{23}^2=t, \quad p_{123}^2=M^2.
\end{equation} 
This integral contributes to the one-loop amplitude for $gg \to H g$, and it
was computed for the first time in Ref.~\cite{Baur:1989cm}.   We will see that dimensional regularization will allow us to cancel ultraviolet divergences associated with collinear subtractions, and will facilitate the calculation of the physical mass-dependence of the diagram.

In the $m \to 0$ limit we have two soft, $S_2$ and $S_3$, and three collinear, $C_{k_1||p_1}, C_{k_4||p_3}$, and $C_{k_2||p_2}$, singular limits when $m\to 0$. 
Following our standard approach, we subtract first the soft and then
the collinear limits.  We note that after the soft subtractions the numerator of the subtracted diagram vanishes in the limit where the loop momentum is collinear to external momentum $p_2$, $C_{k_2||p_2}$.   Additional subtractions are therefore required only for the regions $C_{k_1||p_1}$ and $C_{k_4||p_3}$.

This procedure yields an integrand that is free of leading-power mass singularities, 
\begin{eqnarray}
\label{eq:JR-def}
J_R &=& \int \frac{d^d k_1}{i \pi^\dhalf} 
\left\{ \frac{1}{B_1 B_2 B_3 B_4}
\left[ 
1 - \frac{B_1}{s} -\frac{B_4}{t}
\right] 
-\frac{(1-\frac{M^2}{t})}{st} \frac{1}{B_1 B_2}  \frac{1}{x_1 +(1-x_1) \frac{M^2}{t}}
\right. 
\nonumber \\ 
&& \left. 
-\frac{(1-\frac{M^2}{s})}{st} \frac{1}{B_3 B_4} \frac{1}{x_3 +(1-x_3) \frac{M^2}{s}}
\right\}\, .
\end{eqnarray}
In this expression we have chosen $x_1$ and $x_3$ as the physical fractional momenta carried in the collinear limits,
\begin{equation}
x_1 =  -\, \frac{2 k_1 \cdot \eta_1}{2 p_1 \cdot \eta_1} ,\quad x_3 = \frac{2
  k_3 \cdot \eta_3}{2 p_3 \cdot \eta_3}\, .
\end{equation}
Here the $\eta_i$ are, as usual,  arbitrary light-like vectors that are not collinear to the corresponding
$p_i$.    In Eq.\ (\ref{eq:JR-def}), the collinear subtractions induce ultraviolet divergences in four dimensions, which 
for this discussion we take to be regularized dimensionally.     

At leading order in the mass expansion, we have for $J_R$ given in (\ref{eq:JR-def})
\begin{equation}
\label{eq:J-equal-JR}
J_R = \left. J_R \right|_{m =0} + {\cal O}(m^2).  
\end{equation}
On the right-hand side, we have set the mass to zero in the explicit expression for $J_R$, while retaining all mass dependence on the left-hand side. 
As in the case of the one-loop box, massless collinear counterterms vanish, which we can interpret as a cancellation
of infrared and ultraviolet poles.  Nevertheless, (\ref{eq:J-equal-JR}) still holds for the dimensionally-regulated integrals.   

Having made this observation,  
we can solve (\ref{eq:J-equal-JR}) for original massive integral that defines $J$, Eq.\ (\ref{eq:massive-box}), obtaining
\begin{eqnarray}
\label{eq:1loopoffshell-mass1}
J &=& 
\int \frac{d^d k_1}{i \pi^\dhalf} 
\frac{1}{B_1 B_2 B_3 B_4}
\left[ 
\frac{B_1}{s} +\frac{B_4}{t}
\right] 
\nonumber \\ 
&& 
+\int 
\frac{d^d k_1}{i \pi^\dhalf} 
\left\{
\frac{1}{A_1 A_2 A_3 A_4}
\left[ 
1-\frac{A_1}{s} -\frac{A_4}{t}
\right]   
+\frac{(1-\frac{M^2}{t})}{st} \frac{1}{B_1 B_2} \frac{1}{x_1 +(1-x_1) \frac{M^2}{t}}
\right. 
\nonumber \\ 
&& \left. 
+\frac{(1-\frac{M^2}{s})}{st}  \frac{1}{B_3 B_4 }\frac{1}{x_3 +(1-x_3) \frac{M^2}{s}}
\right\} +{\cal O}(m^2)\, .
\end{eqnarray} 
The first integral of the right-hand consists of the two soft counterterms
that correspond to the one-loop massive triangle, which we examined
above.  The second integral contains the {\it massless} one-loop box 
(which is significantly simpler than its massive analogue) with its
soft singularities subtracted.   In the dimensionally-regulated form, the collinear poles
left in the massless box integral are cancelled by the ultraviolet poles of the two
massive counterterms.   This is not an accident, of course, because the ultraviolet
poles of the massive and massless counterterms must be equal, and the 
ultraviolet pole of the massless counterterm cancels its infrared pole.

All terms on the right-hand side of Eq.~(\ref{eq:1loopoffshell-mass1}) can now be integrated. 
This gives for the original integral, $J$, Eq.\ (\ref{eq:massive-box}), the expression, 
\begin{eqnarray}
\label{eq:massive_box_expansion}
 s t J &=&  
\log^2\left( \frac{s}{m^2}\right)
+\log^2\left( \frac{t}{m^2}\right)
-\log^2\left( \frac{M^2}{m^2}\right)
- \frac{\pi^2}{3}  
\nonumber \\ 
&& \hspace{-1cm}
+2 {\rm Li}_2\left(  \frac{M^2-s}{t}\right)
+2 {\rm Li}_2\left(  \frac{M^2-t}{s}\right)
-2 {\rm Li}_2\left(  \frac{(M^2-s)(M^2-t)}{st}\right) +{\cal O}(m^2)\, .
\nonumber \\ 
&&
\end{eqnarray}
This result 
agrees with the
direct calculation of the same integral, including full mass dependence, in 
Ref.~\cite{Baur:1989cm}~\footnote{There is a typographical error in
  the imaginary part of the right-hand side of 
  Eq. (C.6) in Ref~\cite{Baur:1989cm}, where the last term should read
$2 i \pi \log \left(- \frac{s}{t_1} \right)$, rather than $2 i \pi \log \left(- \frac{m_H^2}{t_1} \right)$}.

\subsection{Two-loop massive diagonal box with two off-shell legs}

\begin{figure}[h]
\begin{center}
\includegraphics[width=0.5\textwidth]{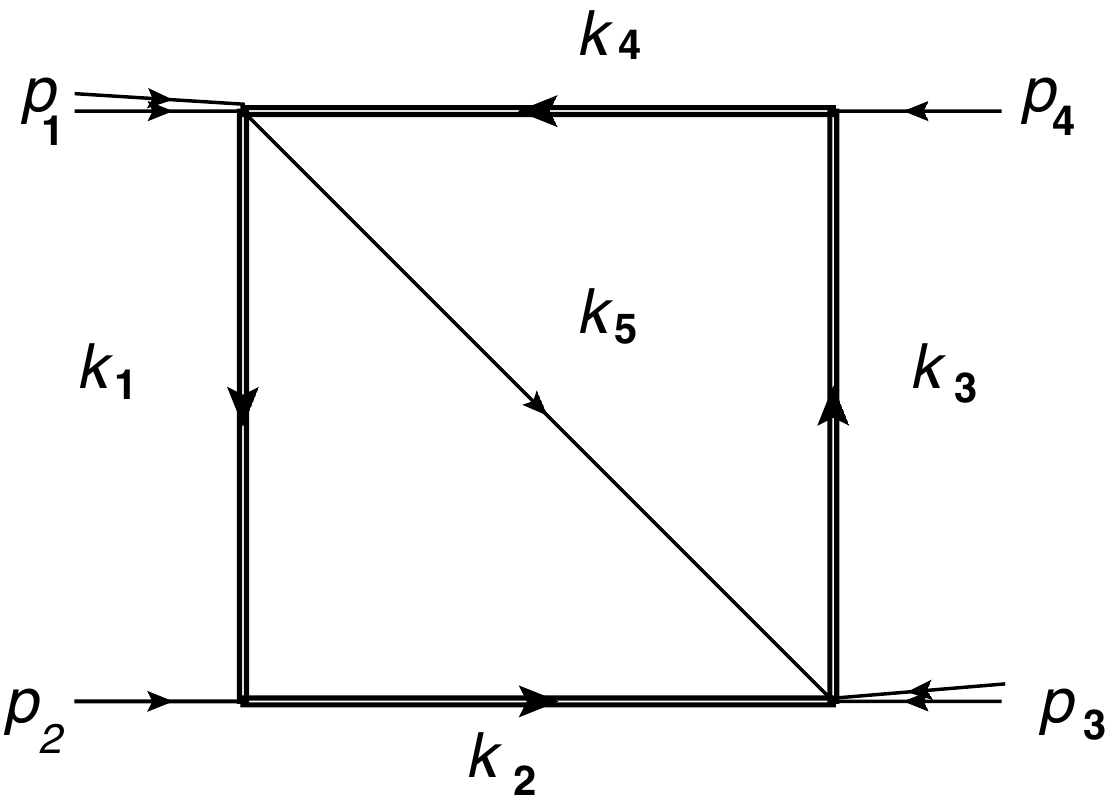}
\caption{\label{fig:massiveDbox} The two-loop diagonal box with
  two off-shell legs. Thick double lines denote massive propagators.}
\end{center}
\end{figure}
As a  final example of our technique for a small mass expansion, we
consider the diagonal box integral with four massive propagators.   Here we will use the
same technique to determine physical mass dependence that we employed in the one-loop case of
the previous subsection, using dimensional regularization to control ultraviolet divergences that
appear in intermediate steps of the calculation.   

The integral we study is,    
\begin{equation}
\label{eq:diagbox_mass_definition}
\DboxM \equiv \int \frac{d^dk}{i \pi^\dhalf}
\frac{d^dl}{i\pi^\dhalf}  \frac{1}{B_1 B_2 B_3 B_4  A_5}\, , \quad 
\end{equation}
which differs from the diagonal box integral, Eq.\ (\ref{eq:diagbox_definition}) only by the introduction of mass $m$ for the 
``outside" lines of the box,
\begin{equation}
B_i = A_i -m^2 + i0, \quad A_i = k_i^2 + i0, \quad i=1 \dots 4\, . 
\end{equation}
The momentum assignments, $k_i$, of the propagators are depicted in
Fig.~\ref{fig:diagonal-box}. One can choose, for example,
\begin{equation}
k_1=l+p_1, \quad k_2=l+p_{12}, \quad  k_3=k+p_{123}, \quad  k_4=k, \quad k_5=k-l\, . 
\end{equation}
The kinematics of the external momenta $p_i$ are given by 
\begin{eqnarray}
  p_2^2 &=& p_4^2 = 0, \quad p_1^2=m_1^2, \quad p_3^2=m_3^2, \nonumber \\[2mm] 
\sum_{i=1}^4 p_i &=& 0,  \quad p_{12}^2=p_{34}^2=s, \quad
p_{23}^2=p_{14}^2=t\, .
\end{eqnarray}
Eq.\ (\ref{eq:diagbox_mass_definition}) is a master integral for the production of one or two Higgs
bosons at hadron colliders.  We have studied the infrared
singularities of the massless diagonal-box in
section~\ref{sec:diagbox-massless}, which becomes divergent in the
double-collinear $C_{1 || 2} C_{4 ||4}$ limit and in the
two single-collinear limits, $C_{1 || 2}$ and $ C_{4 ||4}$. In the massive
diagonal-box that we consider here, these singularities are screened
for finite values of the mass and emerge only as we take the
limit $m \to 0$. Upon integration, these limits  are responsible for 
generating the logarithmic mass singularities $\log^n(m), n=1,2$.
Following steps analogous to those for the massless diagonal box in 
section~\ref{sec:diagbox-massless},
we construct counterterms that remove the mass singularities.  Up to corrections that vanish
with the mass, the fully
subtracted integral is equated to its dimensionally-regulated massless limit, 
\begin{eqnarray}
\label{eq:DboxM-R}
\left. \DboxM\right|_{{\rm R}} &=& 
\int \frac{d^dk_1}{i \pi^\dhalf}
\frac{d^dk_4}{i\pi^\dhalf}  \Bigg\{ 
\frac{1}{B_1 B_2 B_3 B_4  A_5} 
- 
\frac{1}{B_1 B_2} 
\left[ 
\frac{1}{A_3 A_4 A_5}
\right]_{k_1=-x_2 p_2}
\nonumber \\ 
&& 
- 
\frac{1}{B_3 B_4} 
\left[ 
\frac{1}{A_1 A_2 A_5}
\right]_{k_4= x_4 p_4} 
+ 
\frac{1}{B_1 B_2  B_3 B_4} 
\left[ 
\frac{1}{A_5}
\right]_{\begin{array}{l}
{}_{k_4= x_4 p_4,}\\
{}^{ k_1=-x_2 p_2}
\end{array}
}
\Bigg\}
\nonumber \\ 
&=&
\int \frac{d^dk_1}{i \pi^\dhalf}
\frac{d^dk_4}{i\pi^\dhalf}  \Bigg\{ 
\frac{1}{A_1 A_2 A_3 A_4  A_5} 
- 
\frac{1}{A_1 A_2} 
\left[ 
\frac{1}{A_3 A_4 A_5}
\right]_{k_1=-x_2 p_2}
\nonumber \\ 
&& 
- 
\frac{1}{A_3 A_4} 
\left[ 
\frac{1}{A_1 A_2 A_5}
\right]_{k_4= x_4 p_4} 
+ 
\frac{1}{A_1 A_2  A_3 A_4} 
\left[ 
\frac{1}{A_5}
\right]_{\begin{array}{l}
{}_{k_4= x_4 p_4,}\\
{}^{ k_1=-x_2 p_2}
\end{array}
}
\Bigg\} + {\cal O}\left(m^2 \right)\, .
\nonumber \\ &&
\end{eqnarray}
All integrals in the second equality are evaluated with massless
propagators and  all terms except the first integrate to
zero within dimensional regularisation. We can then solve the above
equation for the required massive diagonal-box (up to
${\cal O}(m^2)$). We obtain  
\begin{eqnarray}
  \label{eq:diagbox_massive_from_massless}
&&\int \frac{d^dk_1}{i \pi^\dhalf}
\frac{d^dk_4}{i\pi^\dhalf}  
\frac{1}{B_1 B_2 B_3 B_4  A_5}  
\nonumber \\ 
&\ & \hspace{10mm}   =\
\int \frac{d^dk_1}{i \pi^\dhalf}
\frac{d^dk_4}{i\pi^\dhalf}  \Bigg\{ 
\frac{1}{A_1 A_2 A_3 A_4  A_5} 
 +\frac{1}{B_1 B_2} 
\left[ 
\frac{1}{A_3 A_4 A_5}
\right]_{k_1=-x_2 p_2}
\nonumber \\ 
&& \hspace{20mm}
+
\frac{1}{B_3 B_4} 
\left[ 
\frac{1}{A_1 A_2 A_5}
\right]_{k_4= x_4 p_4} 
- 
\frac{1}{B_1 B_2 B_3 B_4} 
\left[ 
\frac{1}{A_5}
\right]_{\begin{array}{l}
{}_{k_4= x_4 p_4,}\\
{}^{ k_1=-x_2 p_2}
\end{array}} \Bigg\}\, .
\end{eqnarray}
Let us now compare the right-hand side of
Eq.~(\ref{eq:diagbox_massive_from_massless})
and Eq. (\ref{eq:diagb_fin}). We observe that if we set the artificial
scale $\mu$ to the physical mass $\mu=m$ in (\ref{eq:diagb_fin}),
the two expressions differ by terms that integrate to zero within
dimensional regularisation.  We therefore arrive at the following
result for the massive diagonal-box, 
\begin{eqnarray}
\DboxM = \left. \Dbox\right|_{\rm fin}(m) + {\cal O}\left(m^2 \right)  \, ,
 \end{eqnarray}
where $\left. \Dbox\right|_{\rm fin}(m) $ is given by Eq.~(\ref{eq:dbox_fin_polylog}).
We have checked this result against numerical evaluations in the
Euclidean region of results in Ref.~\cite{Bonciani:2016qxi} after
setting $p^2_1=0$ in our expression. We have also checked the coefficients
of the logarithmic expansion (in the $p^2_1=0$ case)  
against the numerical results of an asymptotic expansion 
that is performed using the program of
Refs~\cite{Jantzen:2012mw,Smirnov:2009pb}, based on the
strategy of regions method.

\section{Conclusions}
\label{sec:conclusions}

We have used general properties of the infrared behavior of perturbative amplitudes to
develop a systematic, iterative method for the evaluation of the infrared
singularities in covariant perturbation theory diagrams.   As presented here, the method
applies directly to fixed-angle scattering and particle production of massless and massive particles.
We assumed for this analysis a single hard scale.   The output for each diagram is a finite remainder, plus a sum of 
infrared-sensitive integrals that are generally simpler than the original diagram, and which
depend on fewer external momenta.

The method proceeds through the construction of counterterms that approximate and
cancel divergent behavior of integrands at a local, point-by-point, level.  This approach reflects the 
nested structure of infrared singularities, which extends to all orders. 
Compared to a full ``forest-like" generation of subtractions, we found that an iterative procedure, extending from
regions of lower to higher dimensions, produced a very manageable set of subtractions, even in diagrams with
many singular limits, such as the crossed box.   

   We discussed applications to a number of one- and two-loop diagrams with four external lines.   The method is much more general,
however, and may be particularly useful for isolating infrared poles in dimensionally-regulated multi-particle loop
amplitudes.   

 We have shown that in massless diagrams, the procedure results in a finite remainder, which can in principle
 be evaluated numerically in four dimensions.   It can be applied as well to diagrams with a small
 mass on internal lines, in which case it can be used to derive
 analytic expressions for leading-power mass dependence, including 
 mass-independent terms.   In this context, we studied a number of known cases, and 
 have also derived, and numerically verified, a new result for the two-loop diagonal box diagram.   
  
The procedure outlined in this paper  can be implemented algorithmically, and as such we believe it has 
potential for master integrals beyond two loops, and also for applications to cross sections.
 
\section*{Acknowledgements}

We are grateful to Francesco Moriello who provided numerical checks
for the massive diagonal-box master integral against the results
of~Ref.~\cite{Bonciani:2016qxi}
and the program of Refs~\cite{Jantzen:2012mw,Smirnov:2009pb}.
We thank Z. Capatti, C. Duhr, V. del Duca, F. Dulat, R. Haindl, V. Hirschi,
D. Kermanschach,  S. Lonetti,  Y. Ma,  B. Mistlberger, F. Moriello, A. Ochirov, A. Pelloni, A. Penin, B. Ruijl, Z. Yang
and M. Zeng for many useful discussions or suggestions.  
This research was supported in part by the Pauli Centre for
Theoretical Physics, the National Science Foundation
under Grant PHY-1620628,  by the Swiss National Science Foundation
under contract SNF200021\_179016 and by the European Commission through the ERC grant “pertQCD”.

\newpage
\appendix 

\section{Appendix}
\label{sec:appendix}
\subsection{Polylogarithms}
\label{sec:polylogarithms}
Here, we recall the definition of the generalised Nielsen
polylogarithms that we have used in our article, 
\begin{equation}
S_{np}(x) = \frac{(-1)^{n+p-1}}{(n-1)! p!}\int_0^1 dt
\frac{\log^{n-1}(t) \log^p(1-xt)  }{t}. 
\end{equation}
The $p=1$ value corresponds to standard polylogarithms, 
\begin{equation}
S_{n1}(x) = {\rm Li}_{n+1}(x). 
\end{equation}

\subsection{One-loop master integrals}
\label{sec:collmasters}

For the integration of collinear counterterms for two-loop integrals, 
the massless  one-loop box master integral with one off-shell leg is required. 
This is defined as
\begin{eqnarray}
\label{eq:Boff}
{\rm Boff}(s, t, M) &\equiv& \int \frac{d^dk}{i \pi^\dhalf} \frac{1}{k^2 (k+p_1)^2
              (k+p_{12})^2 (k+p_{123})^2},   
\end{eqnarray}
with 
\begin{equation}
p_1^2=p_2^2=p_3^2=0, \,  p_{123}^2=M^2, \, p_{12}^2=s, \, p_{23}^2=t,
\, u=M^2-s-t.  
\end{equation}
We have
\begin{eqnarray}
{\rm Boff}(s, t, M) &=& \frac{2 (1-2\epsilon)}{s t \epsilon} \left[ 
{\rm Bub}(s)+{\rm Bub}(t)-{\rm Bub}(M^2)
\right] 
+\frac{2}{st} \left( \frac{st}{u}\right)^{-\epsilon}
   \Gamma(1+\epsilon) \Bigg\{
\nonumber \\ 
&& \hspace{-1.3cm} 
{\rm Li}_2 \left( \frac {M-s}{t}\right)
+{\rm Li}_2 \left( \frac {M-t}{s}\right)
-{\rm Li}_2 \left( \frac {(M-s)(M-t)}{st}\right)
-\zeta_2  \nonumber \\ 
&& \hspace{-1.3cm} 
+ \epsilon \left[ 
S_{12} \left( \frac {M-s}{t}\right)
+S_{12} \left( \frac {M-t}{s}\right)
-S_{12} \left( \frac {(M-s)(M-t)}{st}\right)
-\zeta_3
\right]
\nonumber \\ 
&& \hspace{-1.3cm}
+{\cal O}(\epsilon^2)\Bigg\},  
\end{eqnarray}
where the one-loop massless bubble is given by 
\begin{equation}
{\rm Bub}(s) = c_\Gamma \frac{1}{\epsilon (1-2 \epsilon)} (-s)^{-\epsilon}
\end{equation}
and 
\begin{equation}
c_\Gamma = \frac{\Gamma(1+\epsilon) \Gamma^2(1-\epsilon)}{\Gamma(1-2\epsilon)}. 
\end{equation}


\newpage

\end{document}